\documentclass[aps,superscriptaddress,amsmath,amssymb,twocolumn,showpacs,floatfix,english,noshowpacs]{revtex4}

\usepackage{url}
\usepackage{bm,bbm}
\usepackage{graphicx}
\usepackage{color}
\usepackage[colorlinks=true, urlcolor=blue, linkcolor=blue, citecolor=blue, pdftex]{hyperref}
\usepackage[english]{babel}

\usepackage{soul}
\usepackage{blindtext}

\newcommand{\dagga}{{\phantom{\dagger}}}

\begin{document}

\title{Gapless spin liquids in disguise}

\author{Francesco Ferrari}
\affiliation{Institute for Theoretical Physics, Goethe University Frankfurt, Max-von-Laue-Stra{\ss}e 1, D-60438 Frankfurt a.M., Germany}
\author{Alberto Parola}
\affiliation{Dipartimento di Scienza e Alta Tecnologia, Universit\`a dell'Insubria, Via Valleggio 11, I-22100 Como, Italy}
\author{Federico Becca}
\affiliation{Dipartimento di Fisica, Universit\`a di Trieste, Strada Costiera 11, I-34151 Trieste, Italy}

\date{\today}

\begin{abstract}
We show that gapless spin liquids, which are potential candidates to describe the ground state of frustrated Heisenberg models in two dimensions, 
become trivial insulators on cylindrical geometries with an even number of legs. In particular, we report calculations for Gutzwiller-projected 
fermionic states on strips of square and kagome lattices. By choosing different boundary conditions for the fermionic degrees of freedom, both 
gapless and gapped states may be realized, the latter ones having a lower variational energy. The direct evaluation of static and dynamical 
correlation functions, as well as overlaps between different states, allows us to demonstrate the sharp difference between the ground-state 
properties obtained within cylinders or directly in the two-dimensional lattice. Our results shed light on the difficulty to detect {\it bona 
fide} gapless spin liquids in such cylindrical geometries.
\end{abstract}

\maketitle

\section{Introduction}

Quantum spin liquids represent nowadays one of the paradigms of strongly-correlated systems, hosting elementary excitations with fractional
quantum numbers (e.g., $S=1/2$ spinons), emerging gauge fields (e.g., visons or magnetic monopoles), and topological degeneracy of the
ground-state manifold (in case of a gapped spectrum)~\cite{savary2017,zhou2017}. Their unconventional properties are ultimately triggered
by a large entanglement between spins, which remain highly correlated down to zero temperature. In this respect, standard mean-field approaches 
completely fail to give a correct description of spin liquids. In order to overcome this difficulty, elegant {\it escamotages} have been 
introduced, based upon what is now generically known as {\it parton construction}~\cite{baskaran1988,affleck1988,arovas1988,read1989,wen1991}.
Here, the spin degrees of freedom are written in terms of elementary objects (spinons), enlarging the original Hilbert space and introducing 
gauge fields (related to the redundancy of the spinon representation). Mean-field approximations of the spinon Hamiltonian can be performed, 
fixing gauge fields to a given spatial configuration; within fermionic or bosonic representations of spinons, the mean-field approximation 
gives rise to non-interacting models that can be easily handled~\cite{baskaran1988,affleck1988,arovas1988,read1989,wen1991}. Remarkably, in 
some very fortunate cases, the mean-field approximation gives the exact solution of the problem, most notably for the Kitaev compass model 
on the honeycomb lattice~\cite{kitaev2006}. However, for generic spin models, the accuracy of parton-based mean-field methods is questionable. 
Particularly delicate cases arise for gapless spin liquids, which are defined by mean-field theories with a gapless spinon spectrum: here, 
small perturbations (e.g., gauge-field fluctuations) may lead to instabilities towards some spontaneous symmetry breaking, the most celebrated 
one being valence-bond order~\cite{read1989b}. A stable spin liquid (i.e., a state that is not limited to isolated points in the phase diagram) 
can be obtained when the gauge fields have a discrete symmetry and gapped excitations, as in the Kitaev model~\cite{kitaev2006}. Instead, the 
case where the gauge fields have a continuous symmetry and sustain gapless excitations is more problematic. Here, the validity of the 
mean-field approximation has been discussed only in some limiting cases~\cite{hermele2004}, while general implications of low-energy gauge 
fluctuations on the fermionic properties are still unknown. 

One technical aspect that strongly limits the use of mean-field wave functions is the fact that they are defined in the artifically enlarged
Hilbert space. For example, in the fermionic approach for $S=1/2$ models~\cite{wen2002}, the enlarged Hilbert space contains four states per 
site (with zero, one, and two fermions) in contrast to the original spin Hilbert space that is limited to two states per site (up and down 
spins). Therefore, in general, these mean-field wave functions do not give reliable insights into the exact ground-state properties. In order 
to go beyond this approximation, quantum Monte Carlo methods have been very insightful. Here, starting from an auxiliary Hamiltonian for free 
fermions (not necessarily the best mean-field solution), along the Monte Carlo procedure, it is possible to restrict the calculations to the 
original Hilbert space of the spin model by including the Gutzwiller projector~\cite{becca2011}. Then, genuine variational wave functions 
can be defined, thus allowing quantitative predictions for the original spin model. 

The variational approach is particularly suited within the fermionic representation, for which there are polynomial algorithms that allow to 
perform large-size calculations. This approach has been demonstrated insightful to describe $S=1/2$ Heisenberg models on frustrated lattices 
in two spatial dimensions, such as the $J_1-J_2$ models on square~\cite{hu2013,ferrari2020}, triangular~\cite{iqbal2016,ferrari2019}, and 
kagome~\cite{iqbal2013,iqbal2015} lattices. In all these cases, Gutzwiller-projected wave functions have suggested the possibility that a 
{\it gapless} spin liquid may be stabilized in highly-frustrated regimes, the free-fermion spectrum displaying Dirac points. On the honeycomb
lattice the gapless spin liquid does not represent the optimal state, having a slightly higher energy than a state with valence-bond 
order~\cite{ferrari2017}. It is worth mentioning that recent density-matrix renormalization group (DMRG) calculations on the kagome~\cite{he2017} 
and triangular~\cite{hu2019} lattices have given further support to this conclusion. Still, obtaining a gapless spin liquid within DMRG is 
not easy; first of all, because gapped low-entangled states are favored, thus disfavoring highly-entangled gapless phases. In addition, not 
less impactfully, DMRG simulations are usually done on cylindrical geometries, i.e., on $N$-leg ladders with a relatively large number of legs. 
Within this choice of the clusters, the nature of the ground state can be altered with respect to the truly two-dimensional case. Indeed, only 
by using modified boundary conditions along the rungs (namely introducing a fictitious magnetic field piercing the cylinder), it was possible 
to detect gapless points in the kagome and triangular lattices~\cite{he2017,hu2019,zhu2018}. A fictitious magnetic field has been also employed 
in the calculation of the dynamical structure factor for the Heisenberg model on the kagome lattice~\cite{zhu2019}.

Gapless spin liquids represent particularly fragile states, which are expected to be stable only to a limited number of perturbations. In this 
respect, a tiny modification to the Hamiltonian may immediately open a gap in the spectrum or lead to a spinon condensation, giving rise to 
topological order or some sort of spontaneous symmetry breaking. For example, the gapless spin liquid of the Kitaev model, is locally stable 
for small variations of the super-exchange couplings around the isotropic point $J_x=J_y=J_z$, but a magnetic field immediately opens a gap, 
driving the ground state into a topological phase with anyon excitations~\cite{kitaev2006}. In general, considering a ladder geometry, which 
explicitly breaks point-group symmetries, represents also a relevant perturbation that may open a gap.

In this work, we study the fate of the gapless spin-liquid phase obtained on frustrated square and kagome lattice by Gutzwiller-projected 
fermionic wave functions, when cylindrical geometries are considered. We restrict to the case of {\it even} number of legs, in order not to 
introduce further frustration in the system. On cylindrical clusters, either gapless or gapped states can be constructed by playing with the 
boundary conditions of the fermionic operators (still keeping periodic boundary conditions on the physical spins). For the models under 
investigation, the best variational {\it Ansatz} on cylindrical geometries is achieved by states with a gapped excitation spectrum in the 
thermodynamic limit, in striking constrast with the results for isotropic two-dimensional clusters, where the spin liquid states have a gapless 
spectrum and the effect of the fermionic boundary conditions is irrelevant in the thermodynamic limit (the same energy per site is achieved by 
any choice). In addition, the gapped ground state found on cylinders is unique, thus featuring neither spontaneous symmetry breaking nor 
topological degeneracy. It is worth mentioning that this is not forbidden by the Lieb-Schultz-Mattis theorem and its generalizations to 
cylindrical geometries~\cite{lieb1961,affleck1988b}. Our results are relevant for recent studies on both square~\cite{mambrini2006,richter2010,jiang2012b,mezzacapo2012,hu2013,gong2014,doretto2014,morita2015,haghshenas2018,wang2018,liu2018,liao2019,choo2019,ferrari2020,nomura2020,liu2020,hasik2021}
and kagome lattices~\cite{yan2011,depenbrock2012,jiang2012,iqbal2013,clark2013,he2017,liao2017,hering2019}, highlighting the fact that gapless 
spin liquids, as obtained in the truly two-dimensional limit, may turn into trivial gapped phases when constrained into cylindrical geometries.

The paper is organized as follows: in section~\ref{sec:methods}, we outline the parton construction for the Gutzwiller-projected wave function;
in section~\ref{sec:results}, we discuss the results for the square and kagome geometries; finally, in section~\ref{sec:concl}, we draw our
conclusions.

\section{Models and methods}\label{sec:methods}

In the following, we will consider frustrated Heisenberg models
\begin{equation}\label{eq:hamilt}
{\cal H} = \sum_{R,R^\prime} J_{R,R^\prime} {\bf S}_R \cdot {\bf S}_{R^\prime},
\end{equation}
defined on either the square lattice, with primitive vectors ${\bf a}_1=(1,0)$ and ${\bf a}_2=(0,1)$, or the kagome lattice, with ${\bf a}_1=(1,0)$ 
and ${\bf a}_2=(1/2,\sqrt{3}/2)$. For the square lattice, both nearest- ($J_1$) and next-nearest-neighbor ($J_2$) super-exchange couplings are 
considered (with $J_2/J_1=0.5$). For the kagome lattice, only nearest-neighbor ($J$) terms are included. Clusters are identified by two vectors 
${\bf T}_1=L_1 {\bf a}_1$ and ${\bf T}_2=L_2 {\bf a}_2$, with $L_1 \gg L_2$. The finite width $L_2$, which is taken to be even, defines the 
cylindrical geometry; periodic boundary conditions along both directions will be considered, in order not to explicitly break translational 
symmetries.
 
We consider the fermionic approach, in which the $S=1/2$ spin operators are represented by the so-called Abrikosov fermions~\cite{wen2002}:
\begin{equation}\label{eq:Sabrikosov}
{\bf S}_R = \frac{1}{2} \sum_{\alpha,\beta} c_{R,\alpha}^\dagger 
\boldsymbol{\sigma}_{\alpha,\beta} c_{R,\beta}^\dagga.
\end{equation}
Here $c_{R,\alpha}^\dagga$ ($c_{R,\alpha}^\dagger$) destroys (creates) a fermion with spin $\alpha=\uparrow,\downarrow$ on site $R$, and the 
vector $\boldsymbol{\sigma}=(\sigma_x,\sigma_y,\sigma_z)$ is the set of Pauli matrices. The fermionic representation enlarges the local Hilbert
space, including unphysical configurations with zero and two fermions per site. 

The construction of spin-liquid wave functions goes as follows. First of all, an auxiliary (fermionic) Hamiltonian is defined:
\begin{equation}\label{eq:generic_mf}
{\cal H}_{0} = \sum_{R,R^\prime} \sum_{\alpha} t_{R,R^\prime} c_{R,\alpha}^\dagger c_{R^\prime,\alpha}^\dagga +
\sum_{R,R^\prime}  \Delta_{R,R^\prime} c_{R,\downarrow}^\dagga c_{R,\uparrow}^\dagga + h.c.,
\end{equation}
where $t_{R,R^\prime}^\dagga=t_{R^\prime,R}^*$ and $\Delta_{R,R^\prime}=\Delta_{R^\prime,R}$ are singlet hopping and pairing terms. Then, the 
unprojected state is obtained by the ground state $|\Phi_0\rangle$ of ${\cal H}_0$. Finally, the variational {\it Ansatz} for the spin model 
is obtained by enforcing the constraint of one fermion per site, thus going back to the original Hilbert space of $S=1/2$ spins. 
This can be achieved by applying the Gutzwiller projector ${\cal P}_G = \prod_R (n_{R,\uparrow} - n_{R,\downarrow})^2$ to $|\Phi_0\rangle$ 
($n_{R,\alpha}=c_{R,\alpha}^\dagger c_{R,\alpha}^\dagga$ being the fermion density on the site $i$):
\begin{equation}\label{eq:finalwf}
|\Psi_0\rangle = {\cal P}_G |\Phi_0\rangle.
\end{equation}
The physical properties of the projected wave function can be assessed by using standard quantum Monte Carlo techniques~\cite{becca2017}.

\begin{figure}
\includegraphics[width=\columnwidth]{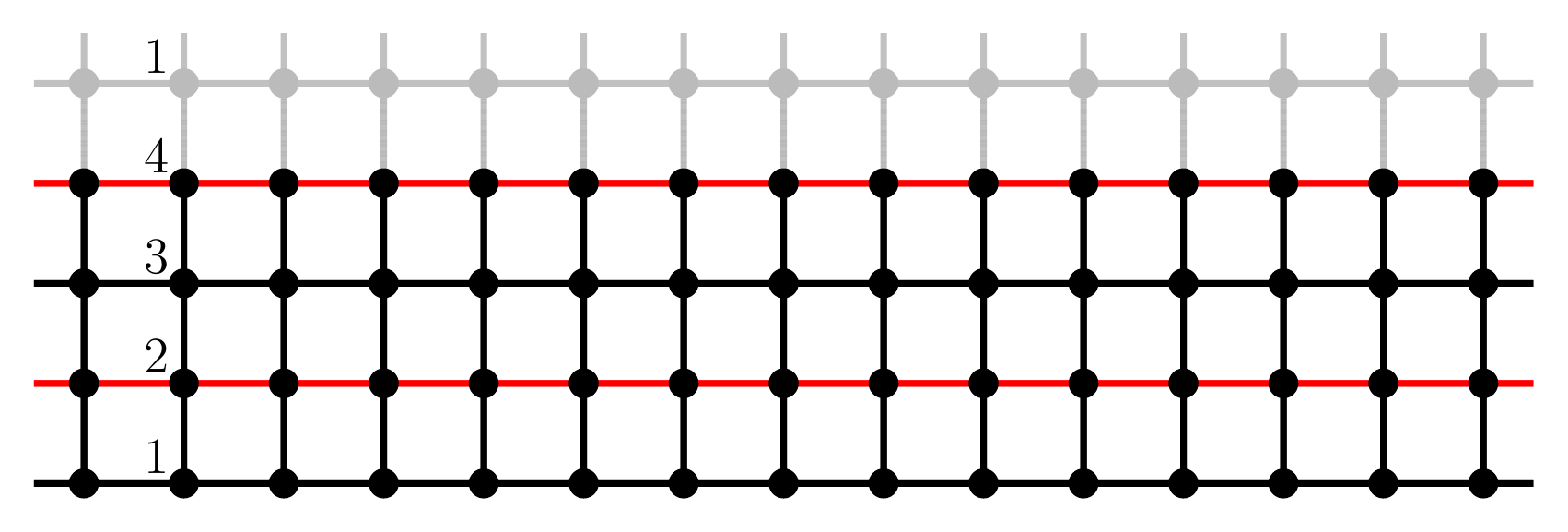}
\includegraphics[width=0.83\columnwidth]{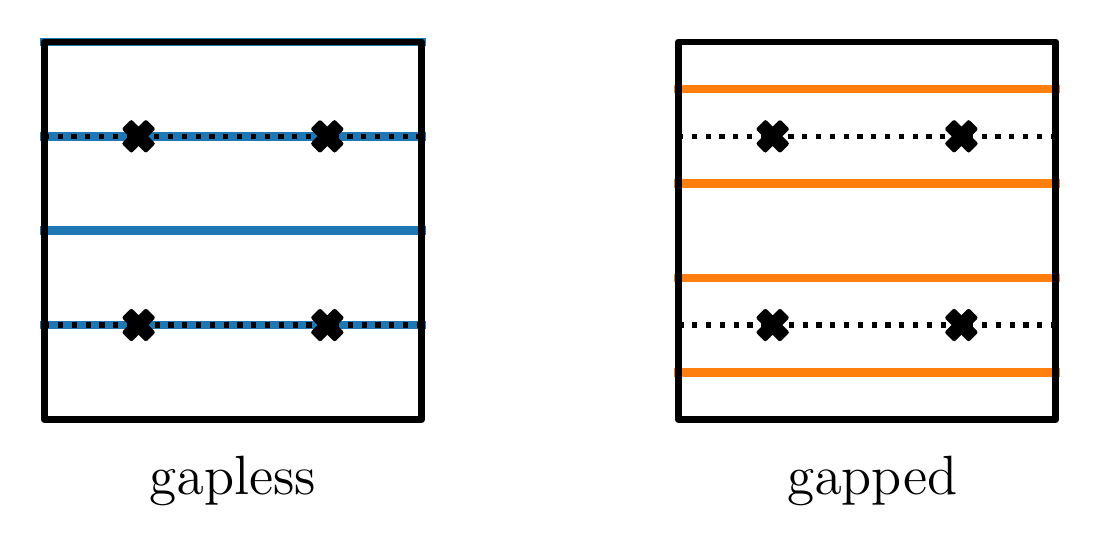}
\caption{\label{fig:square} 
Top panel: schematic illustration of the square lattice cylindrical geometry with $L_2=4$ legs. The site numbering indicates how periodic 
boundary conditions are chosen along the ${\bf T}_2$ direction, namely how the cylinder is wrapped up. Black (red) lines represent positive 
and negative hoppings in the fermionic Hamiltonian~(\ref{eq:generic_mf}) for the $\pi$-flux state~\cite{affleck1988c}. Bottom panels: cuts 
of momenta of the fermionic partons which are allowed by the cylindrical geometry in the top panel. Depending on the choice of the fermionic 
boundary conditions along ${\bf T}_2$, the cuts can hit or avoid the Dirac points in the spinon spectra [${\bf k}=(\pm \pi/2, \pm \pi/2)$, 
indicated by the crosses]. The black square represents the original Brillouin zone, while the dashed lines delimit the reduced Brillouin 
zone of the $\pi$-flux state.}
\end{figure}

\begin{figure}
\includegraphics[width=\columnwidth]{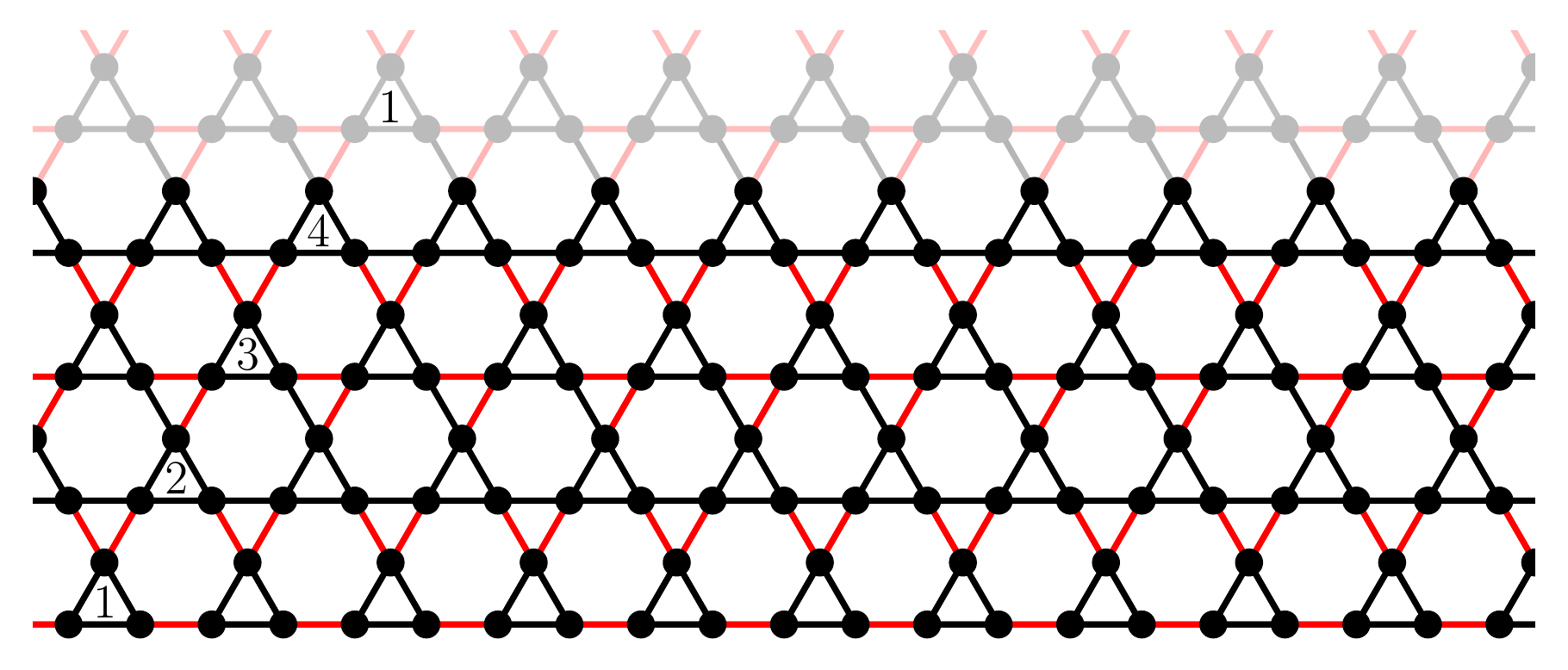}
\includegraphics[width=0.85\columnwidth]{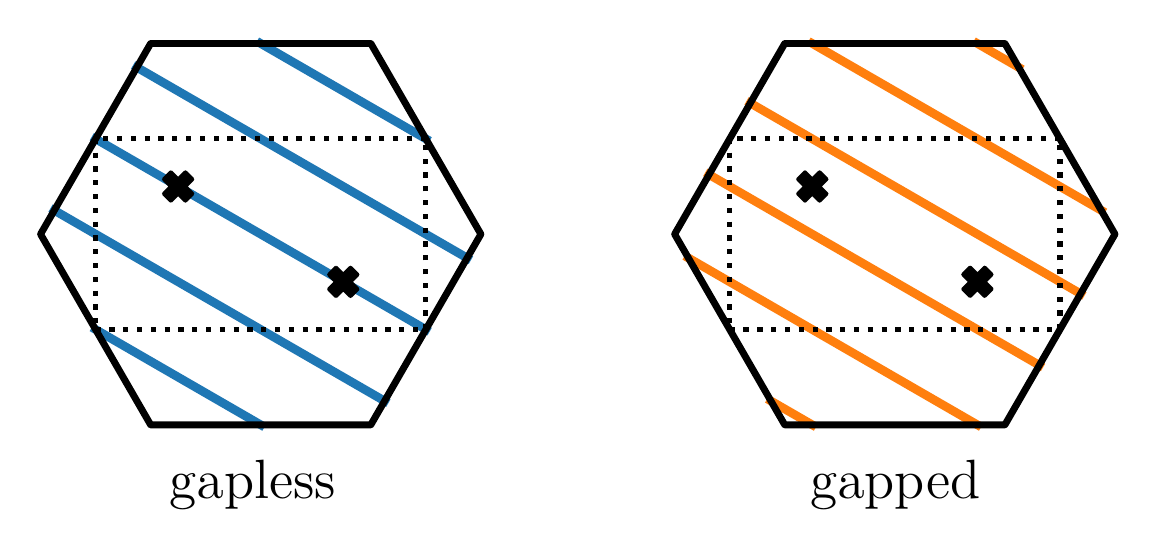}
\caption{\label{fig:kagome}
The same as in Fig.~\ref{fig:square} but for the kagome lattice and the corresponding $\pi$-flux state~\cite{ran2007}.}
\end{figure}

In the following, we will focus on specific spin-liquid {\it Ans\"atze} defined by auxiliary fermionic Hamiltonians which possess a gapless
spectrum with Dirac points in the two-dimensional limit. In particular, we will consider either simple cases with only nearest-neighbor hopping 
parameters, defining non-trivial magnetic fluxes, or cases with both hopping and pairing terms. In the former case, no variational parameters 
are present in the wave function (the hopping parameter defines the energy scale of the auxiliary Hamiltonian); in the latter case, a few 
variational parameters are present and optimized by using standard Monte Carlo schemes~\cite{becca2017}.

\begin{figure*}
\includegraphics[width=\columnwidth]{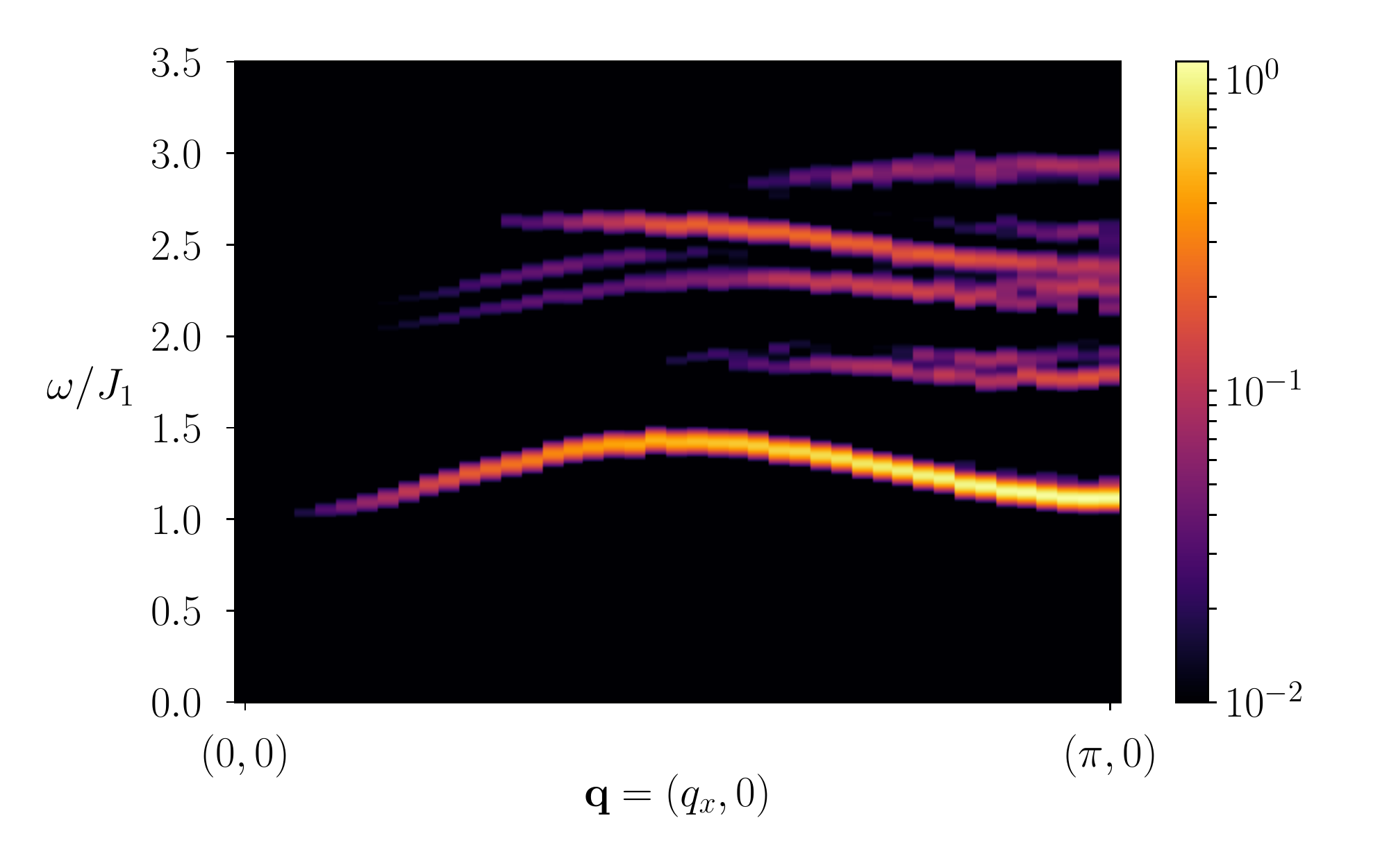}
\includegraphics[width=\columnwidth]{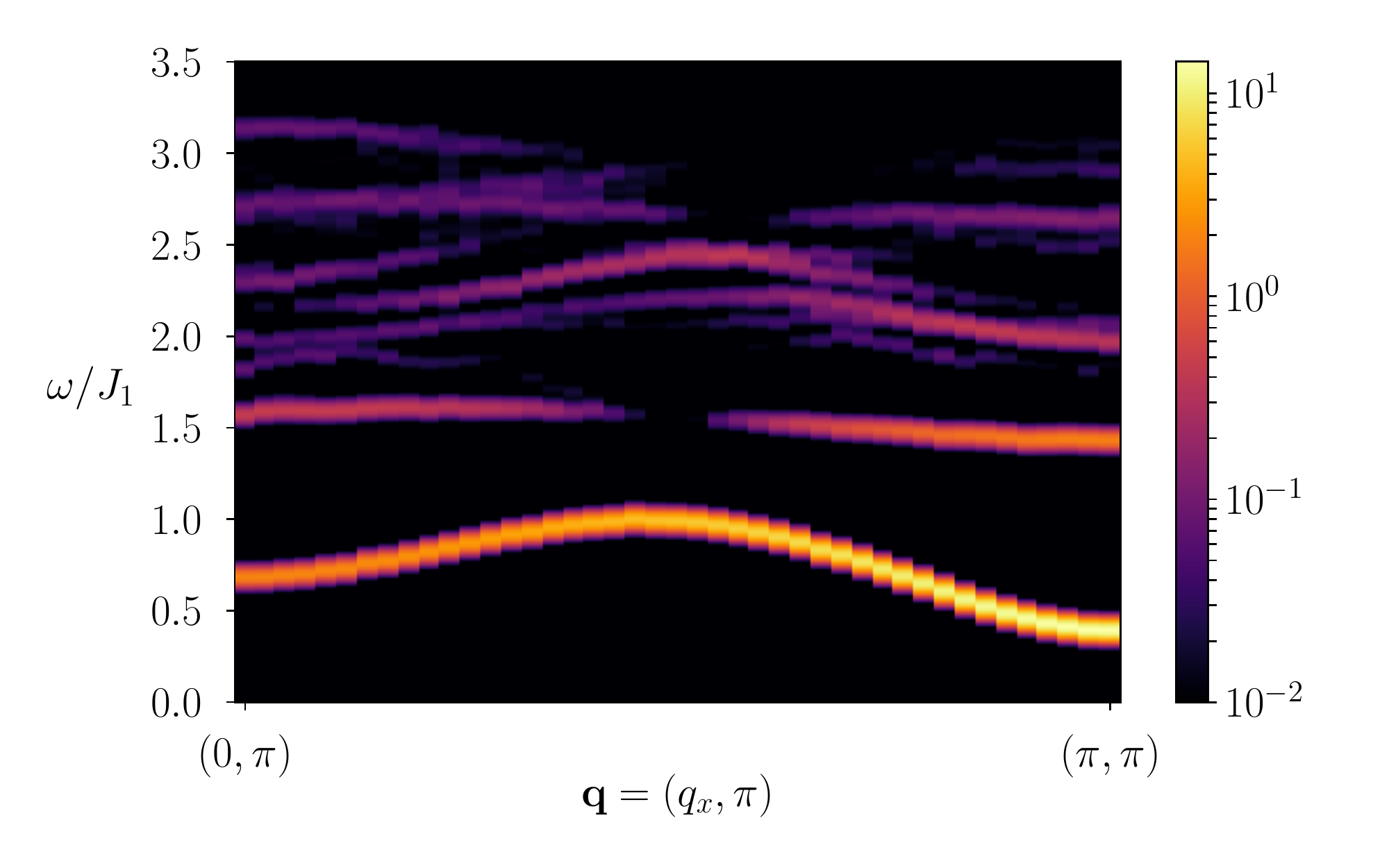}
\includegraphics[width=\columnwidth]{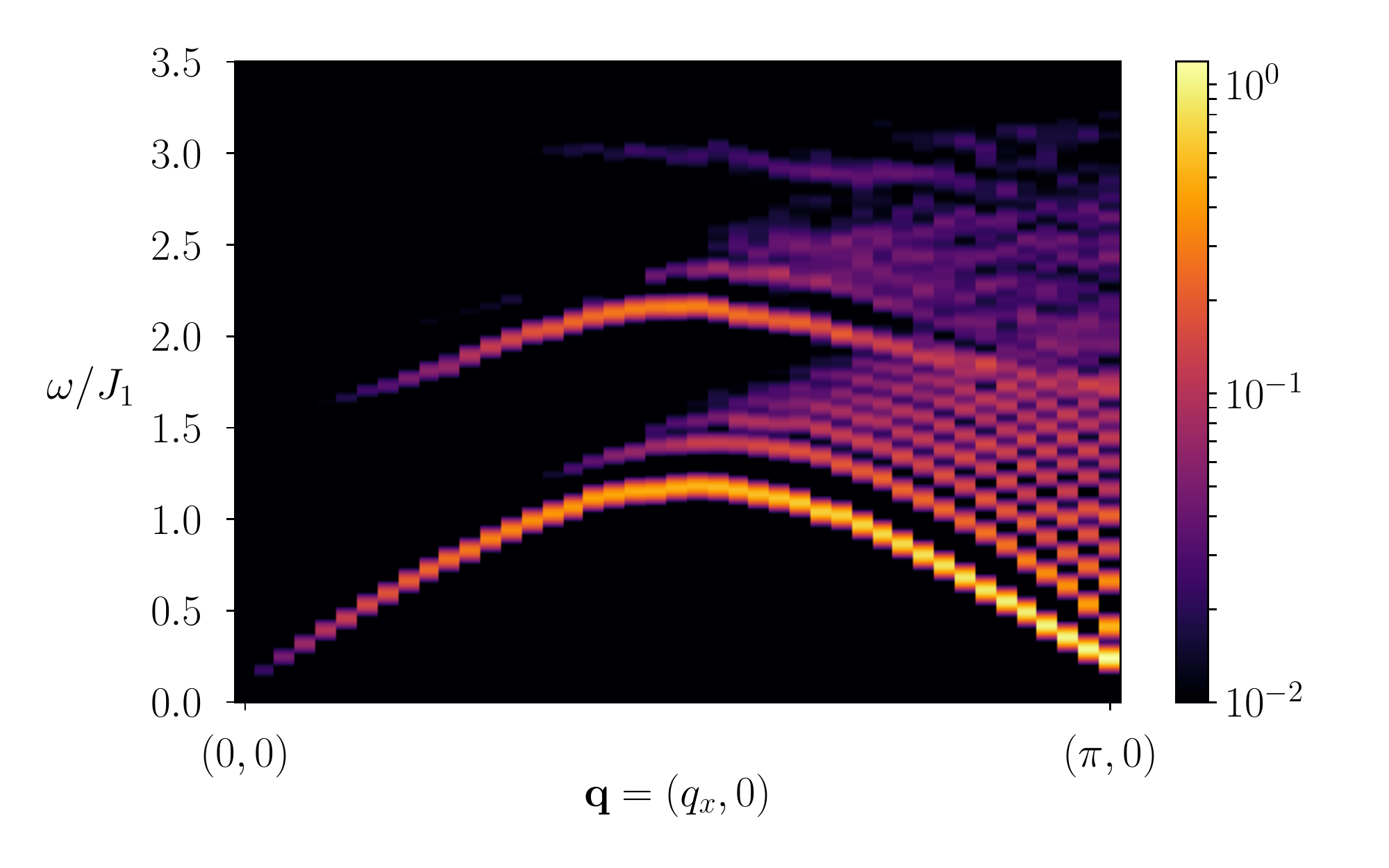}
\includegraphics[width=\columnwidth]{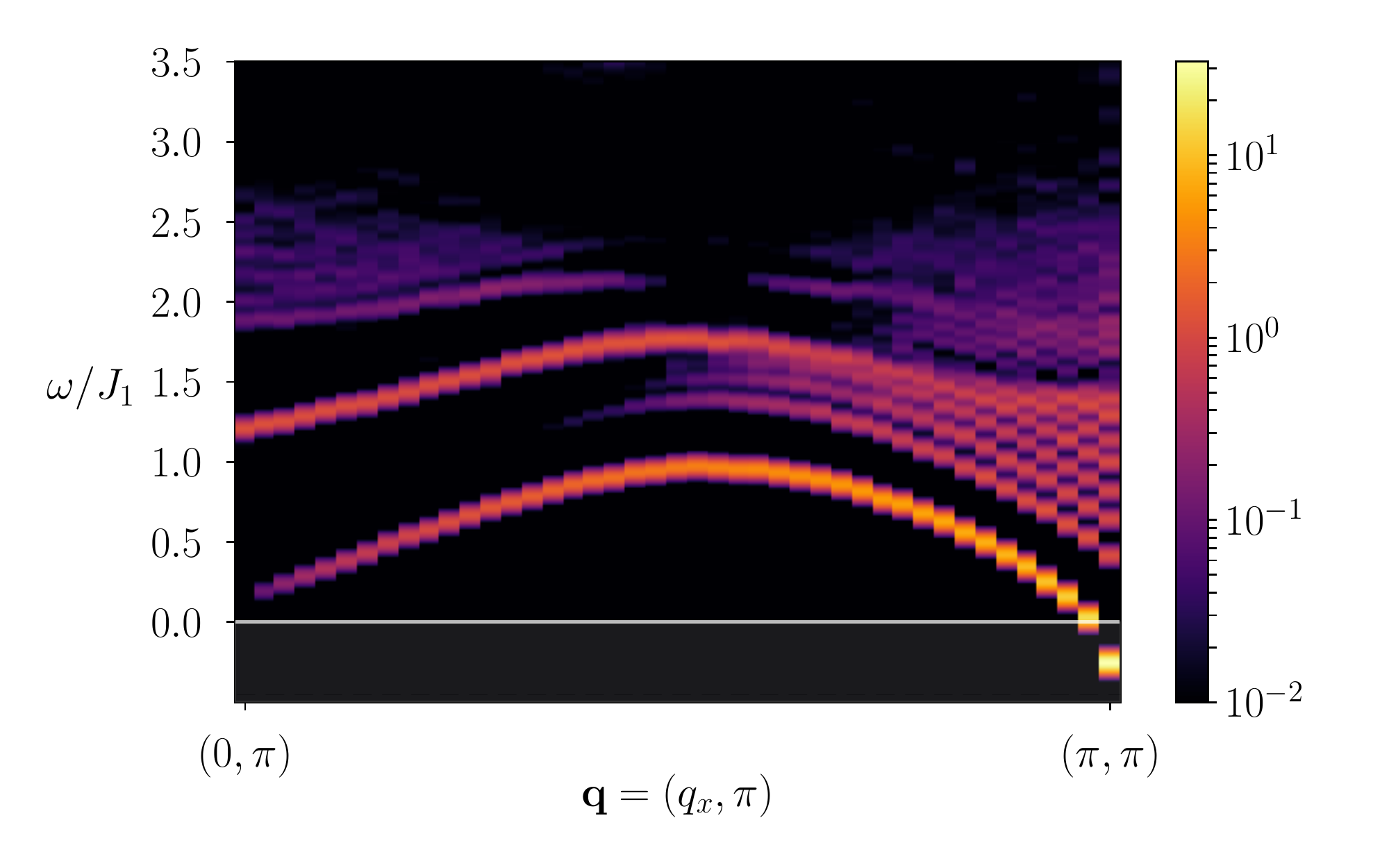}
\caption{\label{fig:square_sqw} 
Dynamical structure factor for the $J_1-J_2$ Heisenberg model on the square lattice with $J_2/J_1=0.5$. The cylindrical geometry with $L_1=84$ 
and $L_2=6$ is taken. The results for the gapped state with PBC-PBC (upper panels) and the gapless one with ABC-ABC (lower panels) are reported for 
two values of the transverse momentum, $q_y=0$ and $q_y=\pi$. A Gaussian smearing of the delta-functions of Eq.~(\ref{eq:dsf}) is applied, with 
$\sigma=0.03J_1$. In the lower-right panel, a negative excitation is observed at ${\bf q}=(\pi,\pi)$. This is an artifact of the numerical method 
caused by the fact that the gapless state is not the optimal variational {\it Ansatz} for the $J_1-J_2$ model on the cylinder (see also the 
discussion in the main text).}
\end{figure*}

Different choices for the boundary conditions on fermionic operators are possible, affecting the free-fermion Hamiltonian~(\ref{eq:generic_mf})
and, consequently, the variational wave function, but not the spin operators~(\ref{eq:Sabrikosov}). In the following, we will restrict ourselves 
to real wave functions, which implies that either periodic boundary conditions (PBC) or anti-periodic boundary conditions (ABC) are possible:
\begin{eqnarray}
c_{R+T_1,\alpha}^\dagga &=& e^{i \theta_1} c_{R,\alpha}^\dagga, \\
c_{R+T_2,\alpha}^\dagga &=& e^{i \theta_2} c_{R,\alpha}^\dagga, 
\end{eqnarray}
where $\theta_1, \theta_2=0(\pi)$ for PBC (ABC).

The important point is that the finiteness of the cylinder width $L_2$ implies a coarse discretization of the momenta along the direction of 
the reciprocal lattice vector ${\bf b}_2$. The discretization is present also in the limit of a infinitely long cylinder, $L_1 \to \infty$, 
where the set of allowed momenta form parallel ``cuts'' running along the ${\bf b}_1$ direction. As a consequence, gapless points in the 
spinon spectrum of the auxiliary Hamiltonian~(\ref{eq:generic_mf}) may be absent on cylinders, if the cuts of momenta do not run through them. 
Therefore, the unprojected spectrum will be gapped also for $L_1 \to \infty$, only because of the finite number of legs $L_2$. As an example, 
let us consider the $\pi$-flux phase on the square lattice, which has Dirac points at ${\bf k}=(\pm \pi/2,\pm \pi/2)$~\cite{affleck1988c}; 
for $L_2=4n$ ($L_2=4n+2$) and ABC (PBC) along ${\bf T}_2$ these k-points are not allowed and the spinon spectrum is gapped, for any value of the 
cylinder length $L_1$. By contrast, for $L_2=4n$ ($L_2=4n+2$) and PBC (ABC) along ${\bf T}_2$, zero-energy modes are present or absent depending 
on the boundary conditions along ${\bf T}_1$: with one choice the gapless points are present for any value of $L_1$, while for the opposite 
choice the spectrum has a finite-size gap, which vanishes in the limit of an infinitely long cylinder $L_1\to \infty$. The case for $L_2=4$ is 
shown in Fig.~\ref{fig:square}. In order to have a unique and well defined variational state, zero-energy modes in Eq.~(\ref{eq:generic_mf}) 
must be avoided on any finite size, thus leaving to only three possibilities of boundary conditions for each choice of $L_1$ and $L_2$: two of 
them will correspond to fully-gapped states, while the third one will correspond to a state that will become gapless for $L_1 \to \infty$ and 
will be dubbed ``gapless'' for simplicity. A similar situation happens in the kagome lattice, for the $\pi$-flux phase defined in 
Ref.~\cite{ran2007}. The case with $L_2=4$ is shown in Fig.~\ref{fig:kagome}. Notice that the unprojected (spinon) spectrum is gauge invariant, 
even though the position of the Dirac points in $k$-space is gauge dependent. Therefore, a given choice of the boundary conditions will give 
rise to a gapped or gapless spectrum, independently on the gauge choice. In both $\pi$-flux phases on square and kagome lattices, the fermionic 
Hamiltonian~(\ref{eq:generic_mf}) can accomodate the magnetic fluxes by breaking the translational symmetry (e.g., the unit cell is doubled 
along ${\bf a}_2$). Nevertheless, after Gutzwiller projection, the wave function~(\ref{eq:finalwf}) is fully translationally 
invariant~\cite{wen2002}.

The properties of these variational states can be assessed by computing equal-time spin-spin and dimer-dimer correlations as a function of 
distance along the long side of the cylinder. The former ones are defined as:
\begin{equation}
{\cal S}({\bf R}) = \langle \mathbf{S}_0 \cdot \mathbf{S}_R \rangle,
\end{equation}
where $\langle \cdots \rangle$ denotes the expectation value over the variational state. Instead, dimer-dimer correlations are given by:
\begin{equation}
{\cal D}({\bf R}) = \langle D_0 D_R \rangle - \langle D_0 \rangle \langle D_R \rangle,
\end{equation}
where $D_R=\mathbf{S}_R \cdot \mathbf{S}_{R+a_1}$. 

Up to now, in the definition of gapped and gapless states, we have made reference to the unprojected spectrum of the fermionic Hamiltonian.
The physical excitations of the spin model may be assessed within the variational method by constructing a set of Gutzwiller-projected states 
for each momentum ${\bf q}$~\cite{li2010,ferrari2018}. For the case of a Bravais lattice (e.g., the square lattice), we define a set of triplet 
states:
\begin{equation}\label{eq:qRstate}
|q,R\rangle =  \mathcal{P}_G 
\frac{1}{\sqrt{N}} \sum_{R^\prime}\sum_{\alpha} e^{i {\bf q} \cdot {\bf R^\prime}} s_{\alpha} 
c^\dagger_{R^\prime+R,\alpha}c^\dagga_{R^\prime,\alpha} |\Phi_0\rangle.
\end{equation}
where $N=L_1L_2$ and $s_{\alpha}=1$ ($s_{\alpha}=-1$) for $\alpha=\uparrow$ ($\alpha=\downarrow$); these states are labelled by $R$, 
which runs over all lattice vectors. Then, the adopted PBC or ABC are imposed to $c^\dagger_{R^\prime+R,\alpha}$, whenever $R^\prime+R$ falls 
outside of the original cluster. Low-energy excited states are given by suitable linear combinations:
\begin{equation}\label{eq:psinq}
|\Psi_n^q\rangle=\sum_R A^{n,q}_R |q,R\rangle.
\end{equation}
The parameters $\{ A^{n,q}_R \}$ are obtained once the spin Hamiltonian of Eq.~(\ref{eq:hamilt}) is fully specified. In fact, for any given 
momentum ${\bf q}$, we can consider the Schr{\"o}dinger equation, restricting the form of its eigenvectors to the one of Eq.~(\ref{eq:psinq}),
i.e., ${ {\cal H}|\Psi_n^q\rangle = E_n^q |\Psi_n^q\rangle }$. Then, expanding everything in terms of $\{|q,R\rangle\}_R$, we get to the 
following generalized eigenvalue problem
\begin{equation}\label{eq:general_eig_prob}
\sum_{R^\prime} \langle q,R|{\cal H}|q,R^\prime \rangle  A^{n,q}_{R^\prime} = E_n^q \sum_{R^\prime} \langle q,R|q,R^\prime \rangle 
A^{n,q}_{R^\prime},
\end{equation}
which is solved to find the expansion coefficients $A^{n,q}_R$ and the energies $E_n^q$ of the excitations. All the matrix elements,
$\langle q,R|{\cal H}|q,R^\prime \rangle$ and $\langle q,R|q,R^\prime \rangle$, are evaluated within the Monte Carlo procedure, by 
sampling according to the variational ground-state wave function~\cite{ferrari2018}. Finally the dynamical structure factor is computed by:
\begin{equation}\label{eq:dsf}
S^{z}({\bf q},\omega) = \sum_n |\langle \Psi_{n}^q | S^{z}_q | \Psi_0 \rangle|^2 \delta(\omega-E_{n}^q+E_0^{\rm var}),
\end{equation}
where $E_0^{\rm var}$ is the variational energy of $|\Psi_0 \rangle$ and 
\begin{equation}
S^{z}_q=\frac{1}{\sqrt{N}} \sum_R e^{i{\bf q} {\bf R}} S^{z}_R.
\end{equation}
In the case of a lattice with a basis of $n_b$ sites (e.g., the kagome lattice with $n_b=3$), the generalization is straighforward, 
with the only modification that particle-hole excitations~(\ref{eq:qRstate}) acquire basis indices, attached to creation and annihilation 
operators. The dynamical structure factor~(\ref{eq:dsf}) becomes a $n_b \times n_b$ matrix, constructed from spin operators on the elements
of the basis (with a phase factor that depends upon the actual spatial position of each sites). Then, the physical (experimentally measurable) 
quantity is the sum of all its elements.

\begin{figure}
\includegraphics[width=\columnwidth]{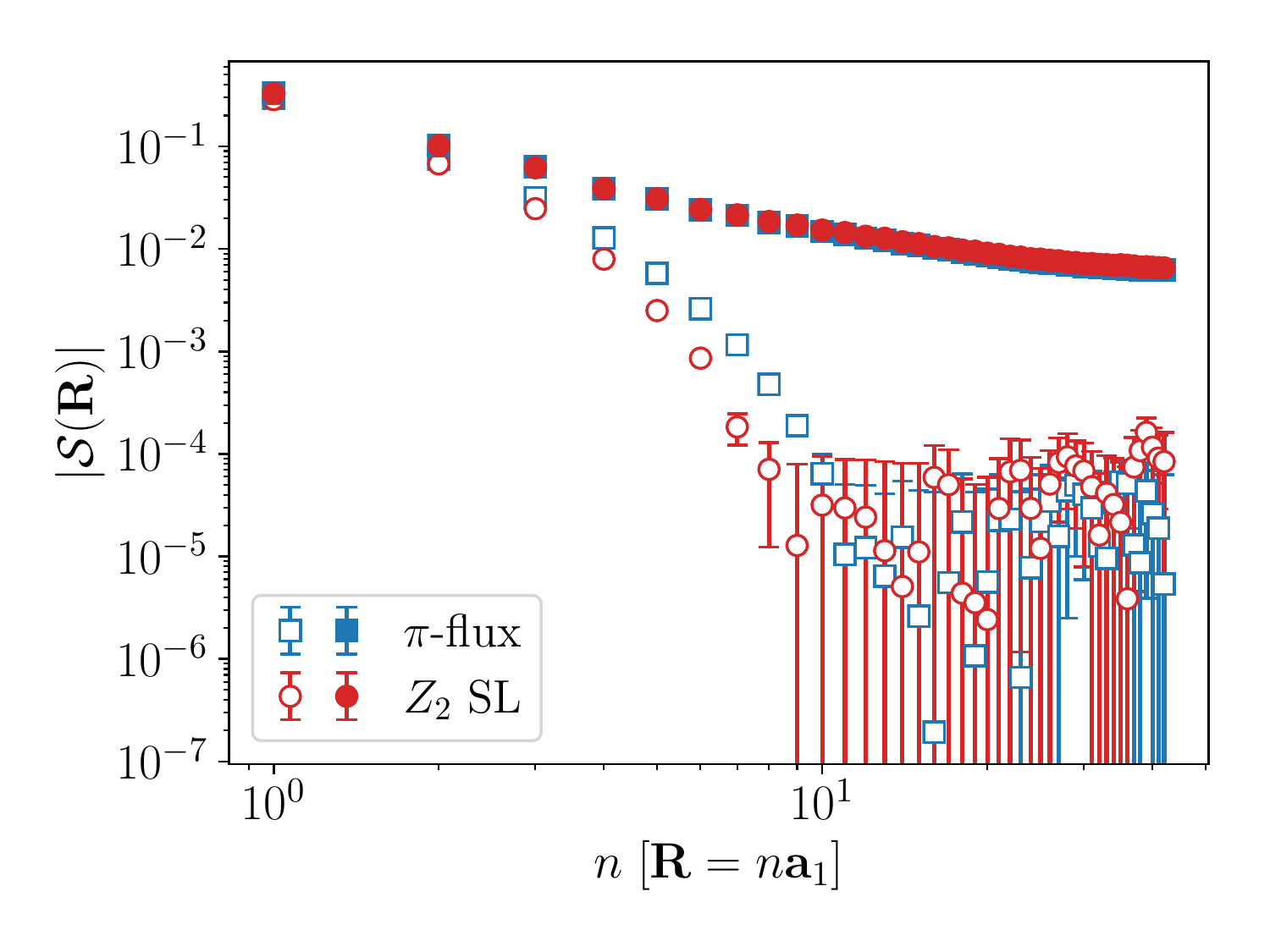}
\includegraphics[width=\columnwidth]{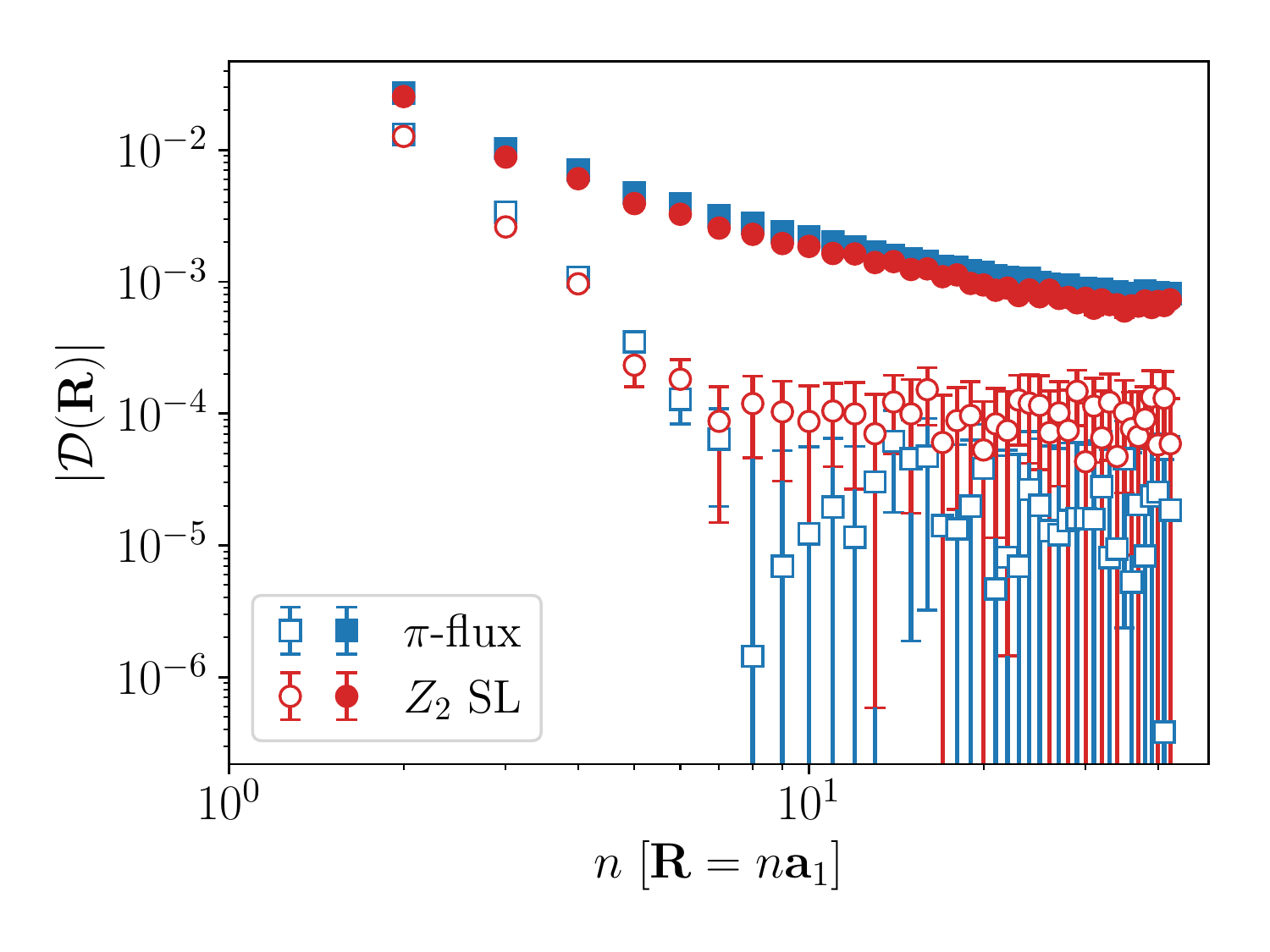}
\caption{\label{fig:piflux_spinspin} 
Comparison of the absolute value of the spin-spin ($|{\cal S}({\bf R})|$, upper panel) and dimer-dimer ($|{\cal D}({\bf R})|$, lower panel) 
correlations for the $\pi$-flux and the optimal spin-liquid wave functions on the square lattice for $J_2/J_1=0.5$. Calculations are done 
on the cylindrical cluster with $84 \times 6$ sites. Empty symbols refer to the gapped state with PBC-PBC, while full symbols refer to the 
gapless state with ABC-ABC.}
\end{figure}

\begin{figure}
\includegraphics[width=\columnwidth]{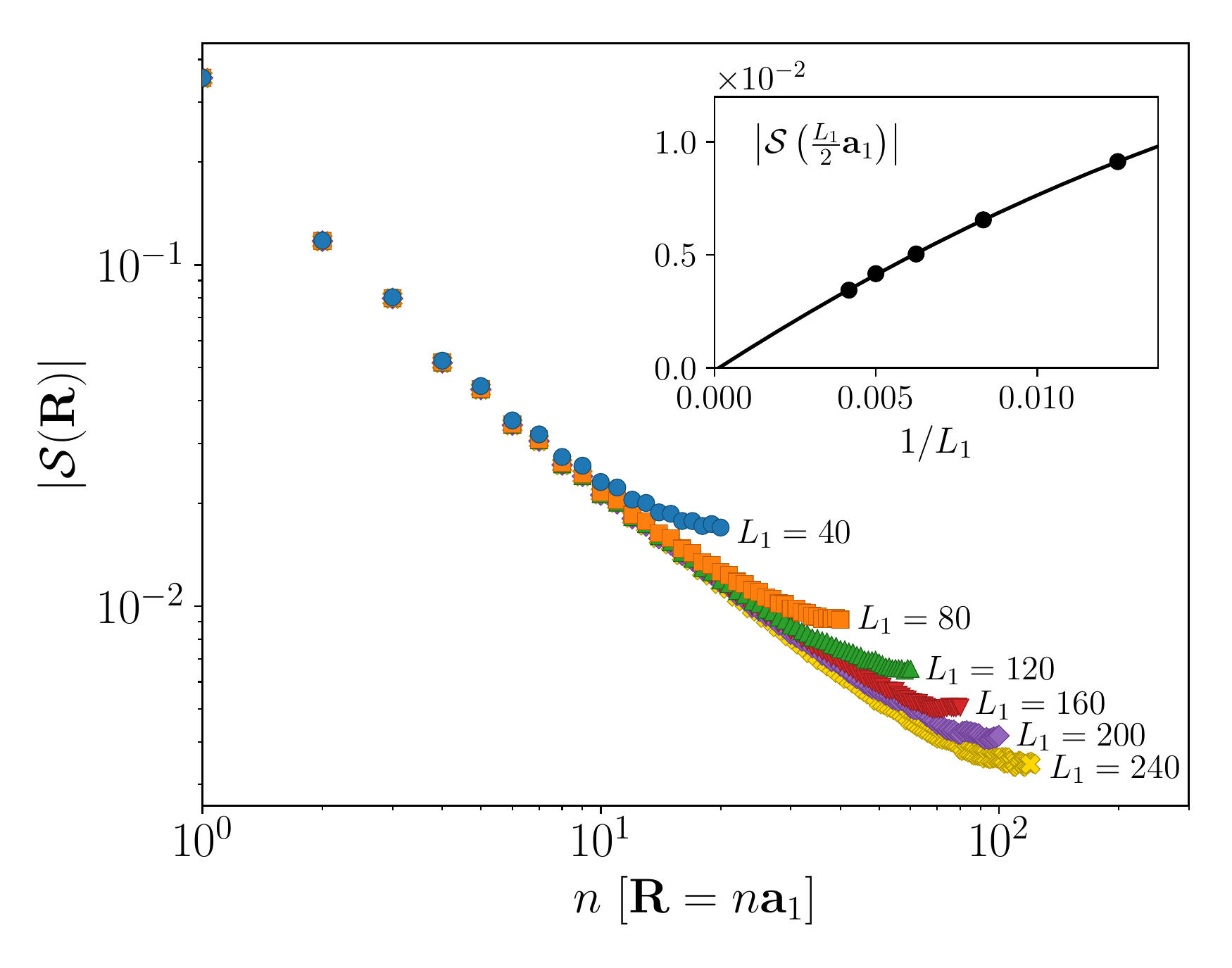}
\caption{\label{fig:piflux_4xL} 
Absolute value of the spin-spin correlations ($|{\cal S}({\bf R})|$) of the gapless $\pi$-flux state (with ABC-PBC) on $L_1 \times 4$ cylinders 
(square lattice). Inset: size scaling of the spin-spin correlations at the maximum distance along ${\bf a}_1$.}
\end{figure}

\section{Results}\label{sec:results}

\subsection{Preliminary considerations on the Kitaev model}
A useful insight on how a gapless spin liquid reveals itself in a finite-size cluster comes from the investigation of the Kitaev compass 
model on the honeycomb lattice~\cite{kitaev2006}. The model can be mapped to free Majorana fermions in a static magnetic field, where fluxes 
piercing the hexagons are either $0$ or $\pi$. This auxiliary model is defined in an extended Hilbert space and the physical states are 
obtained by a projection. Most importantly, all the physical energies of the spin model also belong to the spectrum of the auxiliary Hamiltonian. 
Then, the study of the flux states of free Majorana fermions on the honeycomb lattice provides the ground state energy and the excitation 
spectrum of the Kitaev model. In the thermodynamic limit, the isotropic model ($J_x=J_y=J_z$) displays a gapless spinon branch, due to the 
excitations of Majorana fermions, and a gapped vison spectrum, corresponding to changes in the flux pattern. On the cylindrical clusters 
defined by ${\bf T}_1$ and ${\bf T}_2$ (with $L_1=4n$ and $L_2=2m$), the Lieb theorem~\cite{lieb1994} holds, implying zero flux through all 
hexagonal plaquettes. This choice of the flux pattern leaves four possible orthogonal candidates for the ground state, corresponding to the 
four possible boundary conditions. Zero-energy modes are present only when both $L_1$ and $L_2$ are multiple of $3$. In this case, there are 
two gapped states and one gapless state (for $L_1 \to \infty$), while the fourth one has zero-energy modes. Remarkably, the lowest energy is 
never obtained by the gapless wave function and, therefore, on such cylindrical geometries, the Kitaev model with $J_x=J_y=J_z$ is fully gapped.

\begin{figure}
\includegraphics[width=\columnwidth]{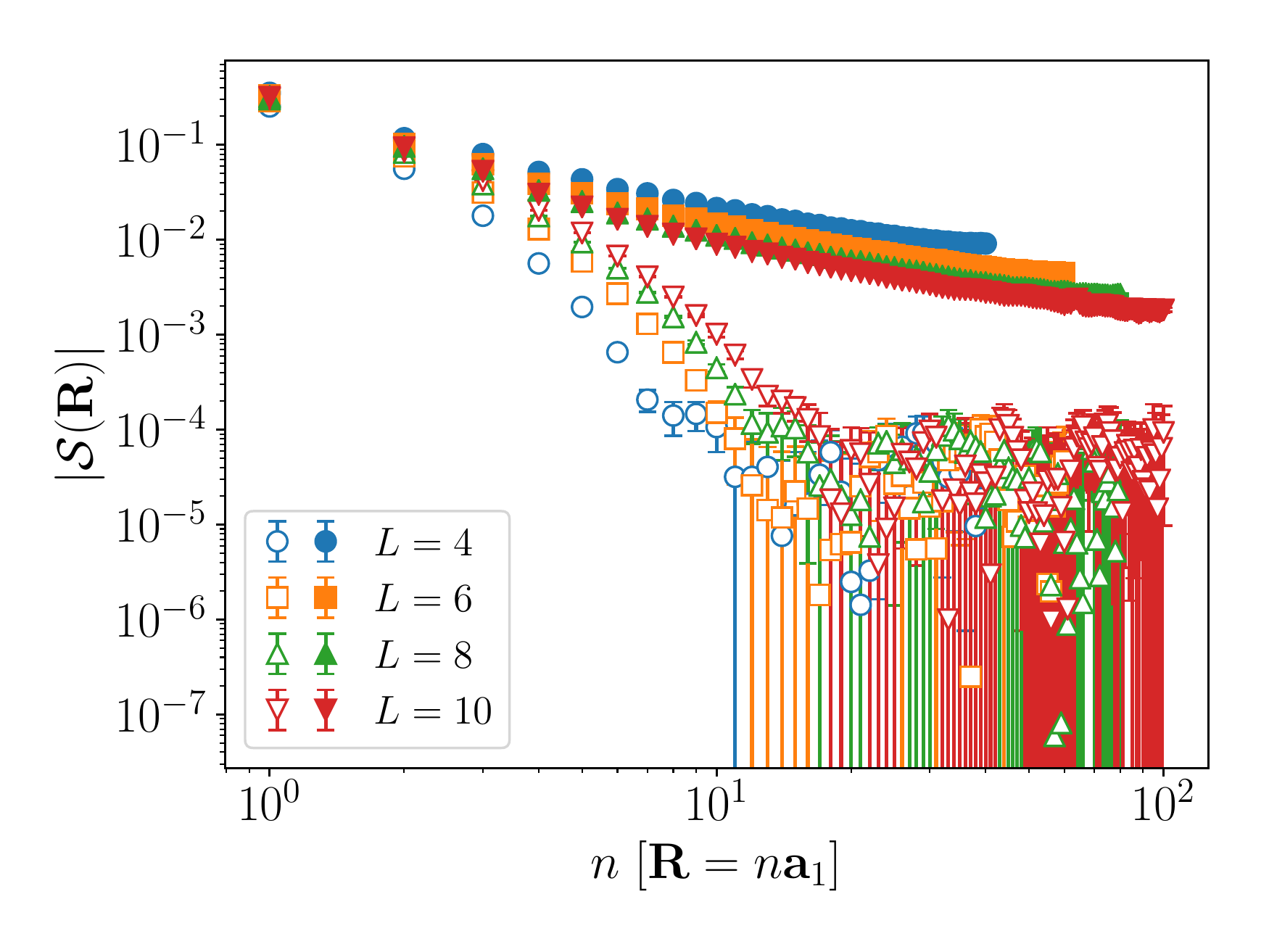}
\includegraphics[width=\columnwidth]{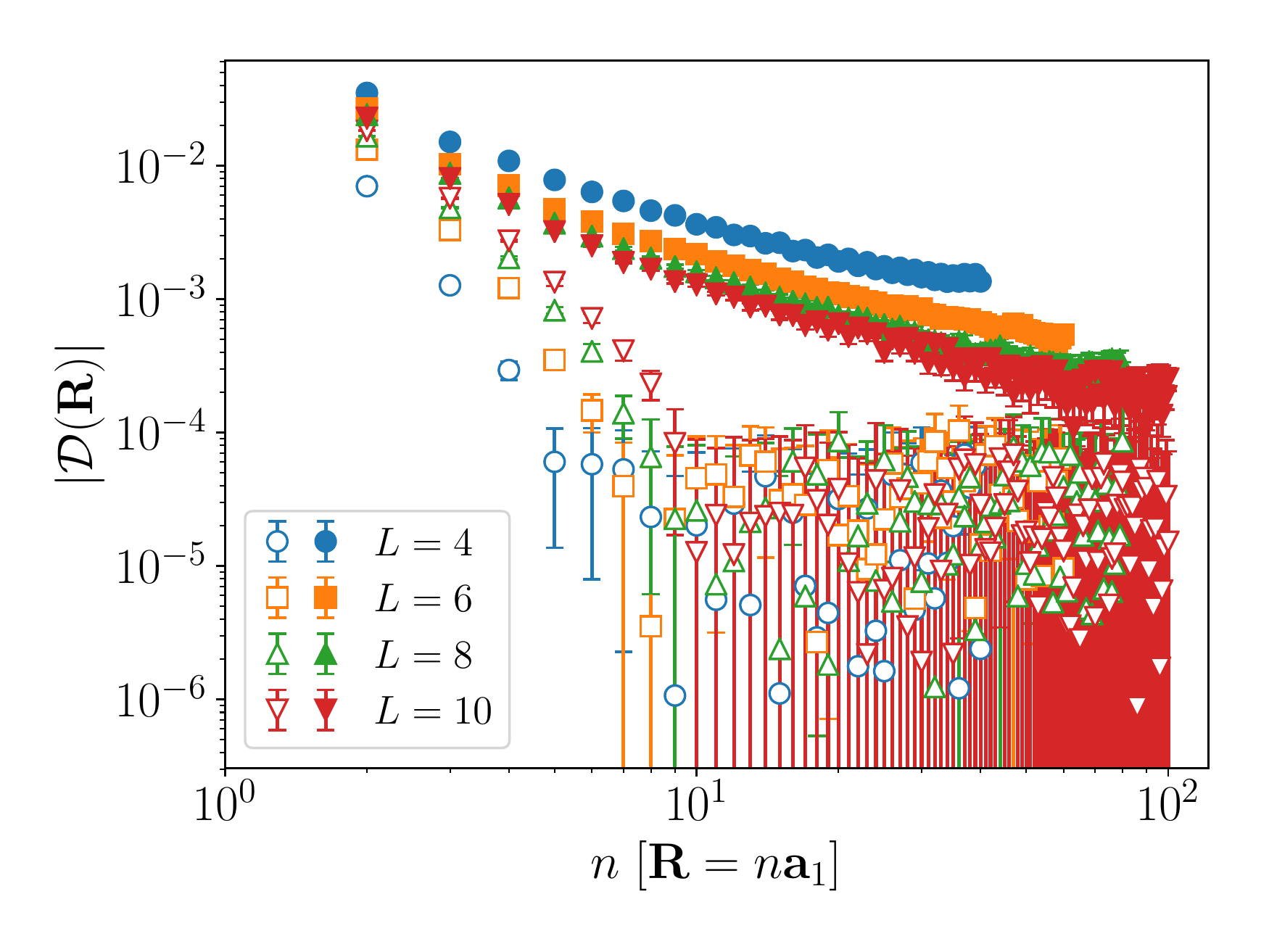}
\caption{\label{fig:largeL_spinspin} 
Absolute value of the spin-spin ($|{\cal S}({\bf R})|$, upper panel) and dimer-dimer ($|{\cal D}({\bf R})|$, lower panel) correlations on 
$20L \times L$ cylinders for the $\pi$-flux state on the square lattice. Empty symbols refer to gapped states (with ABC-ABC for $L=4n$ and 
ABC-PBC for $L=4n+2$), while full symbols refer to gapless states (with ABC-PBC for $L=4n$ and ABC-ABC for $L=4n+2$).}
\end{figure}

\begin{figure}
\includegraphics[width=\columnwidth]{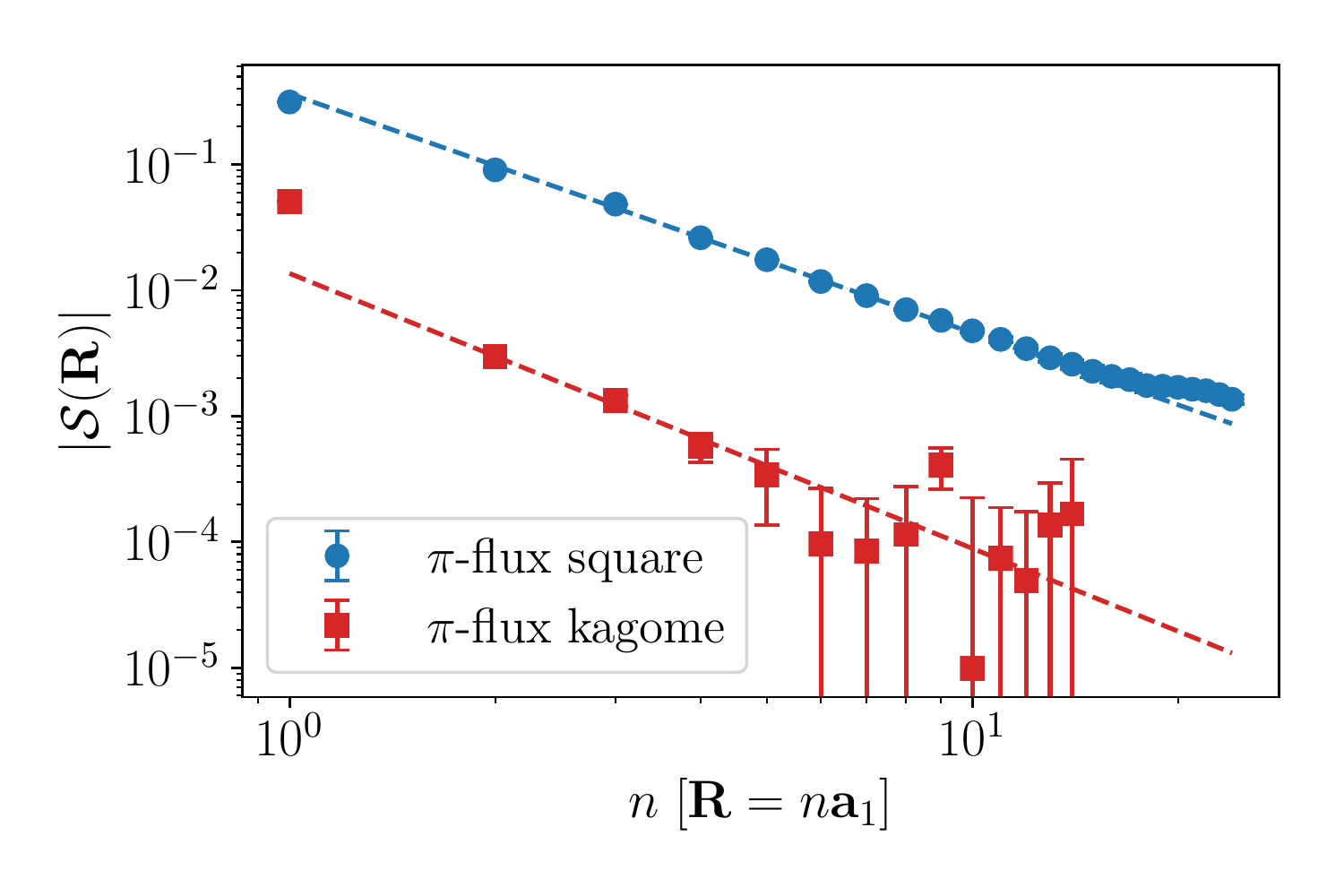}
\caption{\label{fig:squakag}
Absolute value of the spin-spin correlations ($|{\cal S}({\bf R})|$) on the isotropic clusters for the square ($48 \times 48$) and kagome 
($28 \times 28$) lattices. In both cases the simple $\pi$-flux states are considered, with ABC-ABC. Dashed lines are linear fits, compatible 
with a power-law behavior with an exponent $\beta \approx 2$.}
\end{figure}

\subsection{Square-lattice geometry}
Here, we present the results for the square-lattice geometry, see Fig.~\ref{fig:square}, and the spin Hamiltonian~\eqref{eq:hamilt} with 
${J_2/J_1=0.5}$. In the two-dimensional limit, the Gutzwiller-projected wave function suggested the existence of a gapless spin-liquid 
phase~\cite{hu2013,ferrari2020}. The best variational {\it Ansatz} within this class of states has, in addition to uniform nearest-neighbor 
hopping, pairing terms with $d_{x^2-y^2}$ symmetry (at first neighbors, connected by ${\bf a}_1$ and ${\bf a}_2$) and $d_{xy}$ symmetry (at 
fifth neighbors, connected by $2{\bf a}_1 \pm 2{\bf a}_2$). While the point-group symmetry of the hopping and pairing couplings is determined 
by the projective symmetry group of the spin liquid {\it Ansatz}~\cite{wen2002}, the actual values of the parameters are fully optimized by 
the numerical minimization of the variational energy. We note that the simpler case with only hopping and nearest-neighbor pairing corresponds 
to a $U(1)$ spin liquid (gauge equivalent to the the staggered flux state~\cite{affleck1988c}), while the addition of $d_{xy}$ pairing terms 
leads to a $Z_2$ spin liquid~\cite{wen2002}. In both cases, the Hamiltonian~(\ref{eq:generic_mf}) possesses Dirac points at 
${\bf k}=(\pm \pi/2, \pm \pi/2)$. Variational Monte Carlo calculations of the dynamical structure factors suggested that Dirac points survive 
the Gutzwiller projection and are a genuine feature of the low-energy spectrum~\cite{ferrari2018b}. It is worth noting that the simple $\pi$-flux 
state discussed previously, despite having a slightly higher energy than the $Z_2$ spin liquid, gives a quite accurate approximation of the best 
variational {\it Ansatz}.

Let us start by considering a cylinder with $L_1=84$ and $L_2=6$; the large aspect ratio ensures that results are indistinguishable from the 
limit $L_1 \to \infty$. For this cluster, the fermionic Hamiltonian~(\ref{eq:generic_mf}) is gapped by taking PBC along ${\bf T}_2$ and either 
PBC or ABC along ${\bf T}_1$. Instead, the gapless wave function corresponds to ABC along ${\bf T}_1$ and ${\bf T}_2$. The remaining option,
with ABC along ${\bf T}_2$ and PBC along ${\bf T}_1$, is not considered because it implies zero-energy modes in the unprojected spectrum. 

\begin{figure*}
\includegraphics[width=\columnwidth]{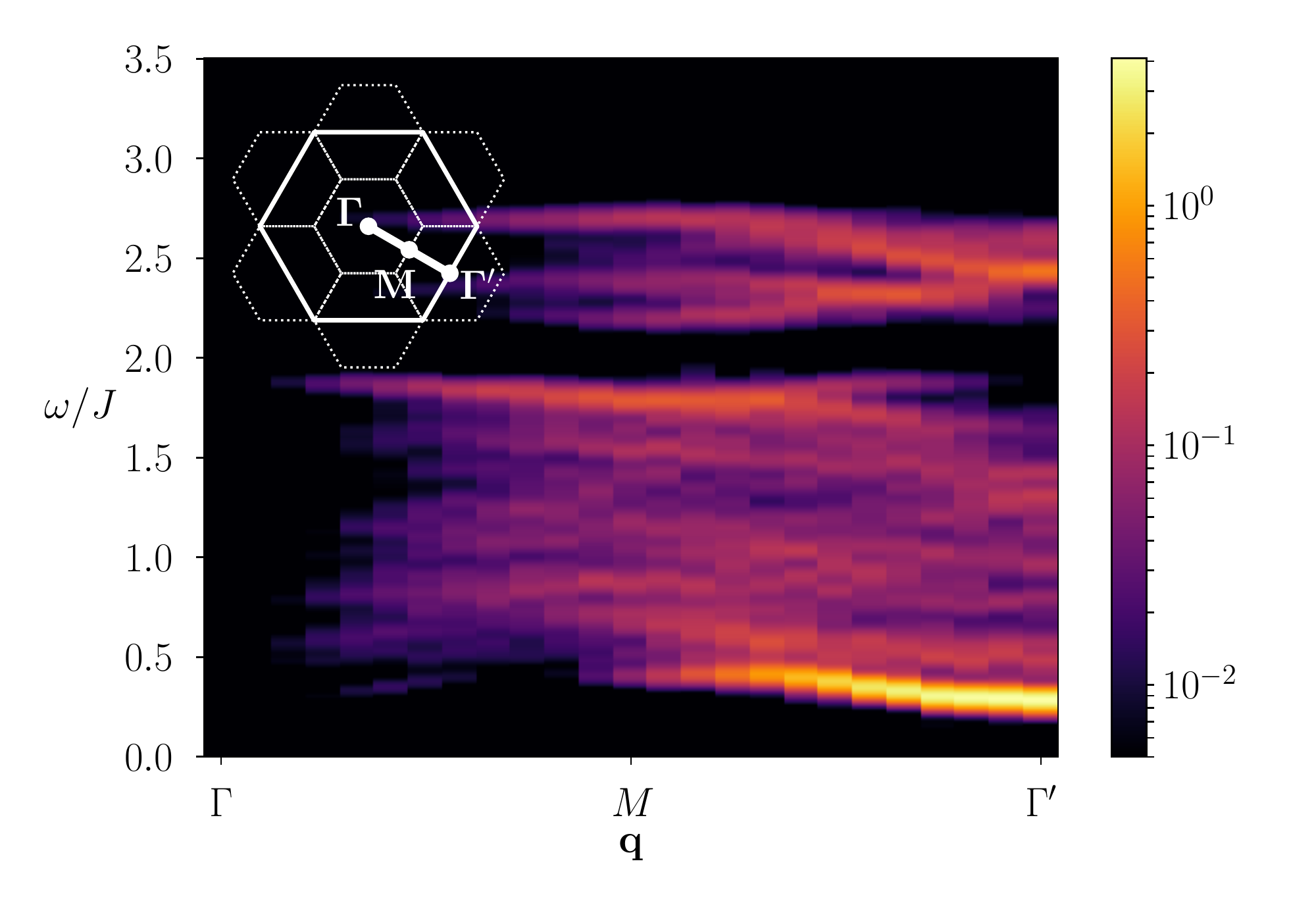}
\includegraphics[width=\columnwidth]{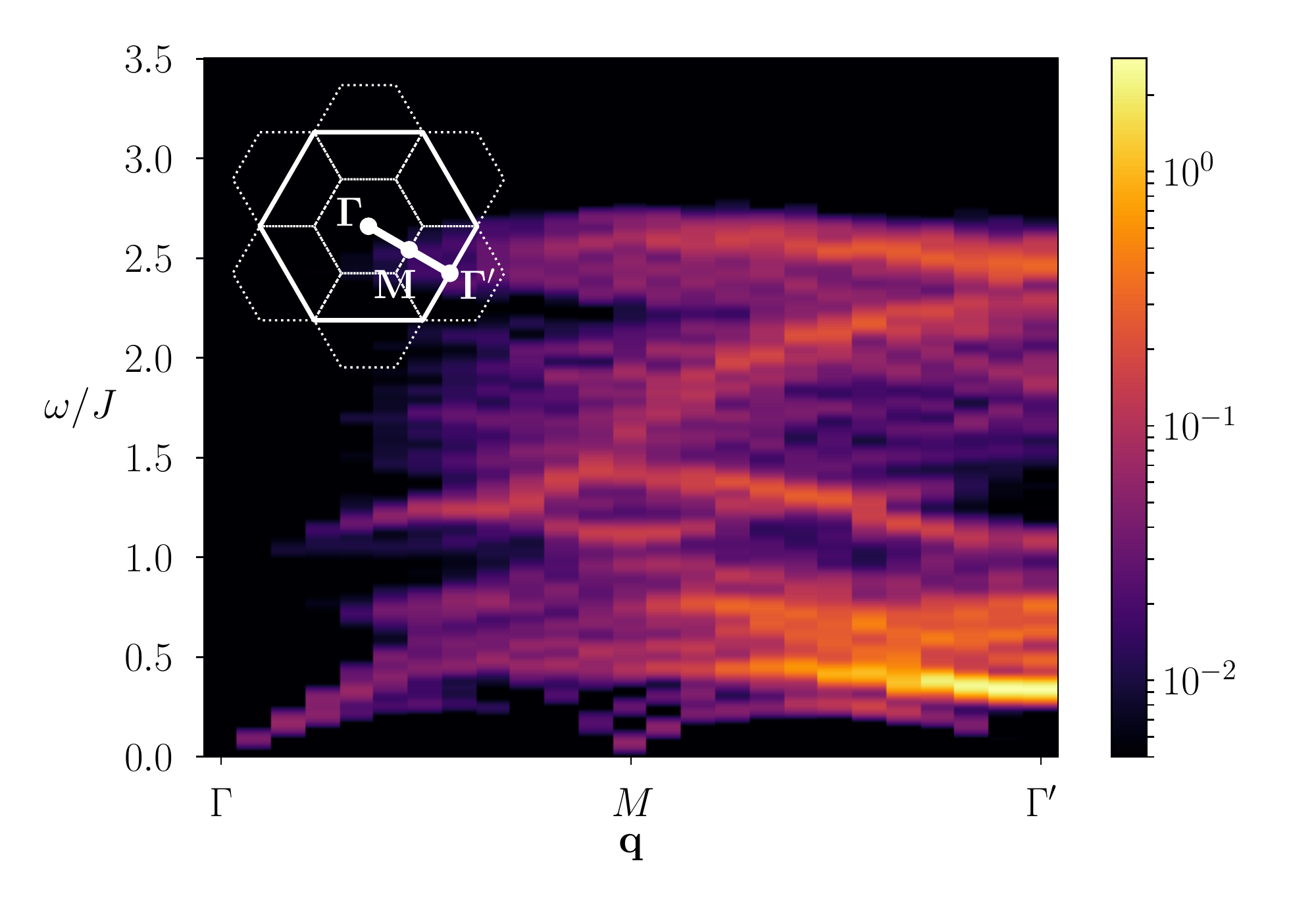}
\caption{\label{fig:kagome_sqw} 
Dynamical structure factor for the Heisenberg model on the kagome lattice. The cylindrical geometry with $L_1=24$ and $L_2=4$ is taken. 
The results for the gapped state with ABC-ABC (left panel) and the gapless one with ABC-PBC (right panel) are reported along the path in 
the extended Brillouin zone indicated in the inset. A Gaussian smearing of the delta-functions of Eq.~(\ref{eq:dsf}) is applied, with 
$\sigma=0.03J$.}
\end{figure*}

The best variational energy, ${E/J_1=-0.49837(1)}$, is obtained by the two gapped states (with PBC-PBC and ABC-PBC), which yield equivalent 
results within the statistical errors. Indeed, these two gapped {\it Ans\"atze} correspond to the same physical state, as demonstrated by the 
fact that the overlap between them (directly computed within Monte Carlo sampling) gives $1$ for the $84 \times 6$ cylinder. On the other hand, 
the gapless {\it Ansatz} has a considerably higher energy, i.e., ${E/J_1=-0.48881(2)}$. Size effects, due to the finiteness of $L_1$, are 
negligible. This represents the first important result of this work, showing that a gapless two-dimensional spin liquid may turn into gapped 
when confined in a cylindrical geometry. The fact that the best variational state actually describes a gapped phase is proven by the dynamical 
structure factor of Eq.~(\ref{eq:dsf}), see Fig.~\ref{fig:square_sqw}. The gapped state shows an intense and dispersive branch, which has a 
minimum at ${\bf q}=(\pi,\pi)$. A weak continuum is also visible at higher energies. By contrast, the gapless state sustains low-energy 
excitations and even a few states with negative energies are present close to ${\bf q}=(\pi,\pi)$. Within this approach, negative excitation 
energies are possible when the reference state $|\Phi_0\rangle$ is not the optimal variational minimum.
Then, it may happen that some of the states defined in Eq.~(\ref{eq:psinq}) end up having a lower energy than the one of $|\Psi_0\rangle$. 
A similar situation was detected in the unfrustrated Heisenberg model by using a spin-liquid state as variational 
{\it Ansatz}~\cite{dallapiazza2015}. However, it is worth stressing that all the wave functions~(\ref{eq:psinq}) constructed from the gapless 
{\it Ansatz} have much higher energies than the one of the gapped state, which remains the best variational approximation for the ground state.

Gapped and gapless states show also remarkably different spin-spin and dimer-dimer correlation functions, see Fig.~\ref{fig:piflux_spinspin}. 
The gapped state possesses rapidly decaying spin-spin and dimer-dimer correlations. The vanishingly small values of the long-range dimer-dimer 
correlations suggests that no valence-bond order is present. Combining this observation with the fact that the two gapped {\it Ans\"atze} 
(with PBC-PBC and ABC-PBC) correspond to the same physical state, we conclude that the ground state on cylindrical geometries is consistent 
with a trivial quantum paramagnet, with neither valence-bond nor topological order.

In order to perform a more detailed analysis of the size-dependence of the correlation functions, we adopt a simpler variational {\it Ansatz}, 
namely the $\pi$-flux state of Fig.~\ref{fig:square}, which contains only hopping terms. The results obtained by the $\pi$-flux state are very 
similar to the ones given by the optimal variational wave function. For example, the energies of gapped and gapless states on the $84 \times 6$ 
cylinder (with $J_2/J_1=0.5$) are $E/J_1=-0.49661(1)$ and $E/J_1=-0.48769(1)$, respectively. Moreover, also static correlation fluctions are 
very similar to those of the optimized $Z_2$ spin liquid, see Fig.~\ref{fig:piflux_spinspin}. Indeed, the overlap between the optimal wave 
function and the simple $\pi$-flux state is approximately $0.8$ ($0.9$) for gapped (gapless) cases. First of all, we prove that the gapless 
{\it Ansatz} has power-law spin-spin correlations, which might be difficult to detect from the results shown in Fig.~\ref{fig:piflux_spinspin}. 
In order to perform a convincing size scaling, we take $L_2=4$ and perform calculations for different values of $L_1$, see Fig.~\ref{fig:piflux_4xL}. 
From these results, we obtain that the spin-spin correlations decay to zero as $1/L_1$ that is fully compatible with a power-law behavior 
$|{\cal S}({\bf R})| \approx 1/|\mathbf{R}|^\beta$ with an exponent $\beta \approx 1$.

Finally, we compute the spin-spin and dimer-dimer correlations of the $\pi$-flux state for different cylindrical geometries with $20L \times L$ 
sites, as shown in Fig.~\ref{fig:largeL_spinspin}. These calculations are easily affordable because the wave function, which does not correspond 
to the optimal state, does not contain any free variational parameter. Thus, no numerical optimization are required. We observe that the 
short-distance correlations of the gapped state strengthen when increasing $L$, rapidly approaching the ones of the gapless state. Still, even 
for cylinders with a relative large width (i.e., $L=10$), the results obtained from gapless and gapped states are quite different from each other, 
indicating a particularly slow convergenece to the truly two-dimensional limit. Interestingly, the spin-spin correlations of the gapless wave 
function show a crossover between a power-law behavior with $\beta \approx 2$ for $|\mathbf{R}| \lesssim L$ and $\beta \approx 1$ for 
$|\mathbf{R}| \gg L$. The former one is typical of the two-dimensional limit, see Fig.~\ref{fig:squakag}, while the latter one is characteristic 
of quasi-one-dimensional systems, as also obtained in Fig.~\ref{fig:piflux_4xL}.

\begin{figure}
\includegraphics[width=\columnwidth]{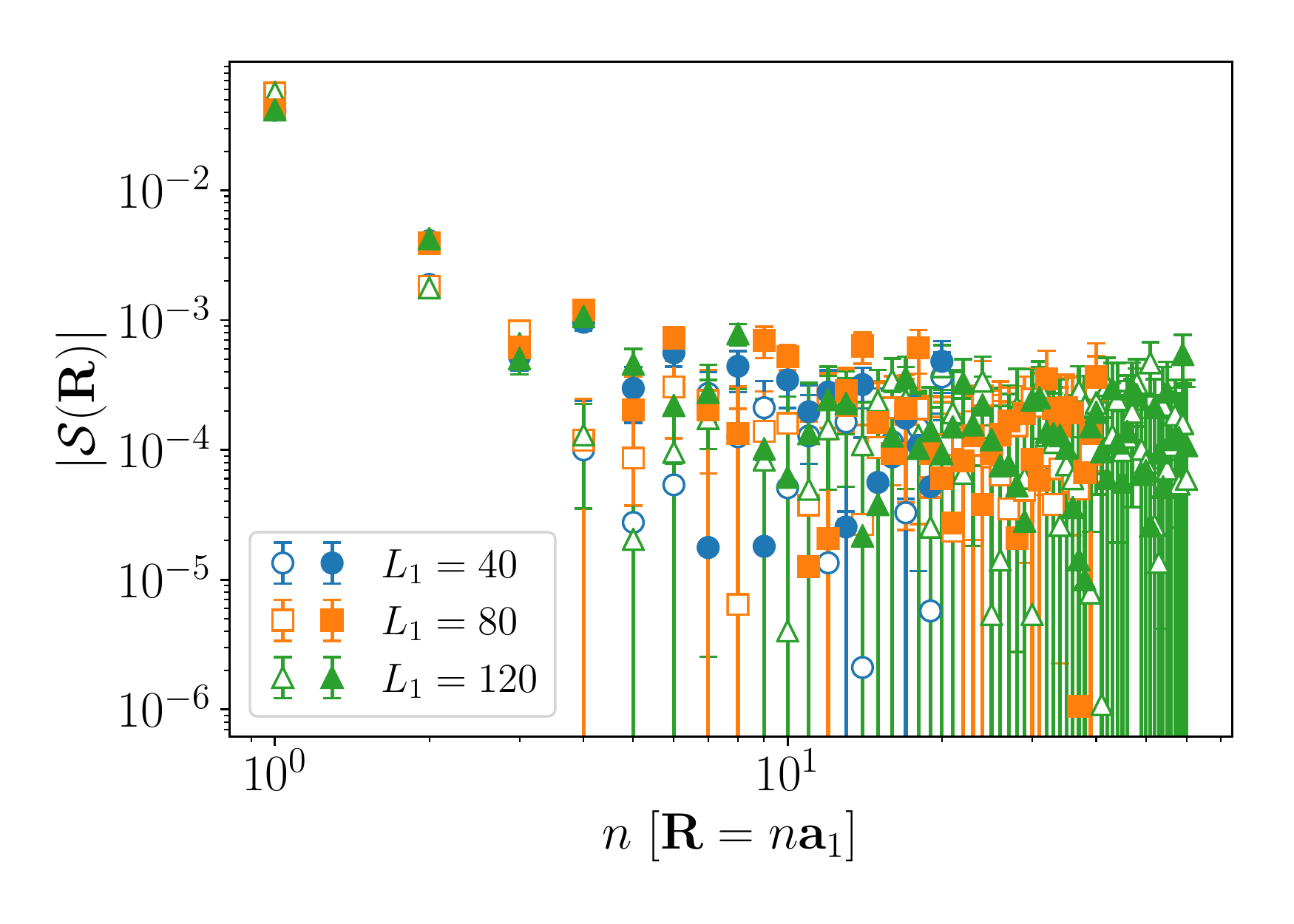}
\caption{\label{fig:kagome_4xL}
Absolute value of the spin-spin correlations ($|{\cal S}({\bf R})|$) on $L_1 \times 4$ cylinders for the $\pi$-flux spin-liquid on the kagome lattice. 
Empty symbols refer to the gapped state with ABC-ABC, while full symbols refer to the gapless state with ABC-PBC.}
\end{figure}

\subsection{Kagome-lattice geometry}
Let us now turn to discuss the results for the kagome-lattice geometry, see Fig.~\ref{fig:kagome}, with nearest-neighbor super-exchange $J$.
Also for this highly-frustrated problem, in the two-dimensional limit, the Gutzwiller-projected fermionic state suggested that the ground
state corresponds to a gapless spin liquid~\cite{ran2007,iqbal2013}. In this case, a very accurate {\it Ansatz} has only nearest-neighbor
hopping, such that fermions experience a $\pi$-flux piercing the elementary unit cell (i.e., $\pi$-flux through hexagons and $0$-flux 
though triangles). As for the $\pi$-flux on the square lattice, the unprojected fermionic spectrum possesses Dirac points~\cite{ran2007}.
This {\it Ansatz} corresponds to a $U(1)$ spin liquid. 

We start by taking a cylinder with $L_1=24$ and $L_2=4$. Here, the fermionic Hamiltonian~(\ref{eq:generic_mf}) is gapped by taking ABC along 
${\bf T}_2$ and either PBC or ABC along ${\bf T}_1$; the gapless wave function has ABC along ${\bf T}_1$ and PBC along ${\bf T}_2$. The fourth
possibility with PBC along both ${\bf T}_1$ and ${\bf T}_2$ leads to zero-energy modes in the spectrum. As for the square lattice, the two
options for the gapped states correspond to the same physical wave function, after Gutzwiller projection (as verified by computing the overlap
between them). 

The gapped state has the best variational energy per site $E/J=-0.43021(3)$, while the gapless wave function has a much higher energy, i.e., 
$E/J=-0.42672(3)$, confirming that also in this case the gapped {\it Ansatz} is favored on the cylindrical geometry, even though the 
two-dimensional case is gapless. In Fig.~\ref{fig:kagome_sqw}, we show the dynamical structure factor for both gapped and gapless states. 
Notice that, in contrast to the square-lattice geometry, here no negative-energy states are obtained for the gapless {\it Ansatz}. In addition, 
the results for this latter case is qualitatively similar to the spectrum recently obtained in the truly two-dimensional case~\cite{zhang2020}. 
A broad continuum of excitations is observed for both the gapped and the gapless state. Notably, in both cases, the maximum of the spectral 
intensity is concentrated around the $\Gamma^\prime$ point (as observed also in Ref.~\cite{zhu2019}), within a weakly dispersive branch of 
excitations which is not separated from the continuum at higher energy.

\begin{figure}
\includegraphics[width=\columnwidth]{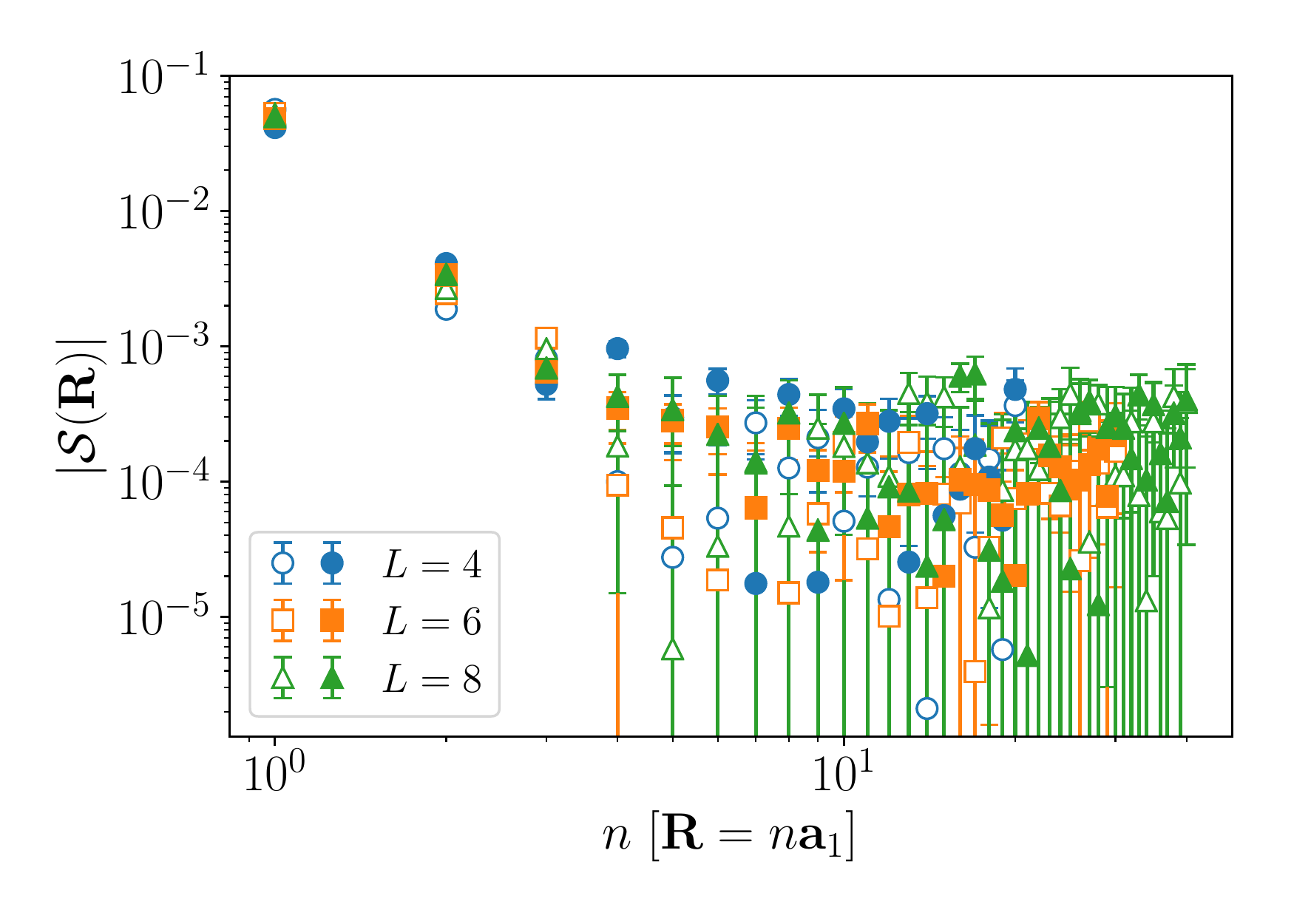}
\caption{\label{fig:kagome_10LxL}
Absolute value of the spin-spin correlations ($|{\cal S}({\bf R})|$) on $10L \times L$ cylinders for the $\pi$-flux spin-liquid on the kagome 
lattice. Empty symbols refer to the gapped state (with ABC-ABC for $L=4n$ and ABC-PBC for $L=4n+2$), while full symbols refer to the gapless 
state (with ABC-PBC for $L=4n$ and ABC-ABC for $L=4n+2$).}
\end{figure}

Another remarkable feature of this {\it Ansatz} on the kagome lattice is the rapid decay of the spin-spin correlation functions, even when 
computed for the gapless wave function. In Fig.~\ref{fig:kagome_4xL}, we report the correlations between spins on the same sublattice (a small 
difference for the three possible sublattices is detected, not changing the qualitative behavior). Here, gapped and gapless {\it Ans\"atze} 
give almost indistinguishable correlation functions for distances larger than $4$ lattice spacings. This aspect is very different from what
we obtained in the square lattice, where gapped and gapless states show rather distinct correlation functions, see Fig~\ref{fig:piflux_spinspin}.
The behavior is not much modified by increasing the number of legs, as reported for $10L \times L$ cylinders in Fig.~\ref{fig:kagome_10LxL}.
Also in this case, there is no apreciable difference between gapped and gapless states, at variance with the square lattice case, where the
gapless state displays sizable correlations, substantially different from the ones of the gapped case. Quite surprisingly, also in the truly 
two-dimensional clusters, spin-spin correlations decay very rapidly with the distance, still being compatible with a power-law decay with an 
exponent $\beta \approx 2$ (similarly to the one obtained in the square lattice), see Fig.~\ref{fig:squakag}. However, in the kagome lattice, 
a small prefactor strongly renormalizes the whole behavior of correlations.

\section{Conclusions}\label{sec:concl}

In conclusions, we performed a systematic study of gapless spin liquids, defined within a fermionic parton construction, on cylindrical geometries. 
We focused on cases that have been widely investigated in the past and have been demonstrated to accurately represent the exact ground-state
properties of two-dimensional square and kagome lattice models~\cite{hu2013,iqbal2013}. In both cases, Dirac points are present in the unprojected 
spectrum of the fermionic spinons. In this work, we considered $L_1 \times L_2$ cylindrical clusters with $L_1 \gg L_2$, in order to include a 
finite-size gap due to the finiteness of the cylinder width $L_2$. By playing with the boundary conditions of fermionic operators, both gapless 
and gapped wave functions can be constructed. Strikingly, the gapped states have a lower variational energy, better approximating the true ground 
state on cylinders. In addition, the gapped states are trivial, i.e., they do not possess local or non-local (topological) order, which is not in 
contrast to the Lieb-Schultz-Mattis theorem ($L_2$ is taken to be even).

Our results are important for two reasons. First of all, they show that two-dimensional gapless spin liquids are quite delicate states of matter
and can be destabilized on constrained geometries, i.e., on cylinders. Here, the same construction that has been used in two dimensions leads
to both gapless and gapped states, just by changing boundary conditions in the underlying parton picture. Gapped states give rise to much lower
variational energies, implying their stabilization on cylinders with an even number of legs. Second, they give a transparent argument to explain 
why previous DMRG calculations~\cite{yan2011,depenbrock2012,jiang2012,jiang2012b,gong2014} failed to detect gapless phases in frustrated models. 
Even though our approach does not give an exact solution of the problem, it strongly suggests that the actual ground state on cylinders is gapped,
thus giving a neat interpretation of DMRG results for the models considered here. In this regard, an important step forward has been the inclusion 
of fictitious magnetic fluxes (in the original spin model), which allowed DMRG to detect the existence of Dirac points in kagome and triangular 
lattices~\cite{he2017,hu2019}.

\section*{Acknowledgements}
We thank D. Poilblanc for discussions in the early stage of the project. F.F. acknowledges support from the Alexander von Humboldt Foundation 
through a postdoctoral Humboldt fellowship.


\begin{thebibliography}{99}

\bibitem{savary2017} L. Savary and L. Balents, Rep. Prog. Phys. {\bf 80}, 016502 (2017).
\bibitem{zhou2017} Y. Zhou, K. Kanoda, and T.-K. Ng, \rmp {\bf 89}, 025003 (2017).
\bibitem{baskaran1988} G. Baskaran and P.W. Anderson, \prb {\bf 37}, 580(R) (1988).
\bibitem{affleck1988} I. Affleck, Z. Zou, T. Hsu, and P.W. Anderson, \prb {\bf 38}, 745 (1988).
\bibitem{arovas1988} D.P. Arovas and A. Auerbach, \prb {\bf 38}, 316 (1988).
\bibitem{read1989} N. Read and S. Sachdev, Nucl. Phys. {\bf B316}, 609 (1989).
\bibitem{wen1991} X.-G. Wen, \prb {\bf 44}, 2664 (1991).
\bibitem{kitaev2006} A. Kitaev, Ann. Phys. {\bf 321}, 2 (2006).
\bibitem{read1989b} N. Read and S. Sachdev, \prl {\bf 62}, 1694 (1989); \prb {\bf 42}, 4568 (1990).
\bibitem{hermele2004} M. Hermele, T. Senthil, M.P.A. Fisher, P.A. Lee, N. Nagaosa, and X.-G. Wen, \prb {\bf 70}, 214437 (2004).
\bibitem{wen2002} X.-G. Wen, \prb {\bf 65}, 165113 (2002).
\bibitem{becca2011} F. Becca, L. Capriotti, A. Parola, and S. Sorella, in {\it Introduction to Frustrated Magnetism: Materials, Experiments, 
Theory} edited by C. Lacroix, P. Mendels, and F. Mila (Springer, 2011), pp. 379-406.
\bibitem{hu2013} W.-J. Hu, F. Becca, A. Parola, and S. Sorella, \prb {\bf 88}, 060402(R) (2013).
\bibitem{ferrari2020} F. Ferrari and F. Becca, \prb {\bf 102}, 014417 (2020).
\bibitem{iqbal2016} Y. Iqbal, W.-J. Hu, R. Thomale, D. Poilblanc, and F. Becca, \prb {\bf 93}, 144411 (2016).
\bibitem{ferrari2019} F. Ferrari and F. Becca, Phys. Rev. X {\bf 9}, 031026 (2019).
\bibitem{iqbal2013} Y. Iqbal, F. Becca, S. Sorella, and D. Poilblanc, \prb {\bf 87}, 060405(R) (2013).
\bibitem{iqbal2015} Y. Iqbal, D. Poilblanc, and F. Becca, \prb {\bf 91}, 020402(R) (2015).
\bibitem{ferrari2017} F. Ferrari, S. Bieri, and F. Becca, \prb {\bf 96}, 104401 (2017).
\bibitem{he2017} Y.-C. He, M.P. Zaletel, M. Oshikawa, and F. Pollmann, Phys. Rev. X {\bf 7}, 031020 (2017).
\bibitem{hu2019} S. Hu, W. Zhu, S. Eggert, and Y.-C. He, \prl {\bf 123}, 207203 (2019).
\bibitem{zhu2018} W. Zhu, X. Chen, Y.-C. He, W. Witczak-Krempa, Science Advances {\bf 4}, 11 (2018).
\bibitem{zhu2019} W. Zhu, S.-S. Gong, and D.N. Sheng, Proc. Natl. Acad. Sci. USA {\bf 116} (12), 5437-5441 (2019). 
\bibitem{lieb1961} E. Lieb, T. Schultz, and D. Mattis, Ann. Phys. {\bf 16}, 407 (1961).
\bibitem{affleck1988b} I. Affleck, \prb {\bf 37}, 5186 (1988).
\bibitem{mambrini2006} M. Mambrini, A. Lauchli, D. Poilblanc, and F. Mila, \prb {\bf 74}, 144422 (2006).
\bibitem{richter2010} J. Richter and J. Schulenburg, Eur. Phys. J. B {\bf 73}, 117 (2010). 
\bibitem{jiang2012b} H.-C. Jiang, H. Yao, and L. Balents, \prb {\bf 86}, 024424 (2012).
\bibitem{mezzacapo2012} F. Mezzacapo, \prb {\bf 86}, 045115 (2012).
\bibitem{gong2014} S.-S. Gong, W. Zhu, D.N. Sheng, O.I. Motrunich, and M.P.A. Fisher, \prl {\bf 113}, 027201 (2014).
\bibitem{doretto2014} R.L. Doretto, \prb {\bf 89}, 104415 (2014).
\bibitem{morita2015} S. Morita, R. Kaneko, and M. Imada, J. Phys. Soc. Jpn. {\bf 84}, 024720 (2015).
\bibitem{haghshenas2018} R. Haghshenas and D.N. Sheng, \prb {\bf 97}, 174408 (2018).
\bibitem{wang2018} L. Wang and A.W. Sandvik, \prl {\bf 121}, 107202 (2018).
\bibitem{liu2018} W.-Y. Liu, S. Dong, C. Wang, Y. Han, H. An, G.-C. Guo, and L. He, \prb {\bf 98}, 241109(R) (2018).
\bibitem{liao2019} H.-J. Liao, J.-G. Liu, L. Wang, and T. Xiang, Phys. Rev. X {\bf 9}, 031041 (2019).
\bibitem{choo2019} K. Choo, T. Neupert, and G. Carleo, \prb {\bf 100}, 125124 (2019).
\bibitem{nomura2020} Y. Nomura, M. Imada, arXiv:2005.14142 (2020).
\bibitem{liu2020} W.-Y. Liu, S.-S. Gong, Y.-B. Li, D. Poilblanc, W.-Q. Chen, Z.-C. Gu, arXiv:2009.01821 (2020).
\bibitem{hasik2021} J. Hasik, D. Poilblanc, F. Becca, SciPost Phys. {\bf 10}, 012 (2021).
\bibitem{yan2011} S. Yan, D. Huse, and S.R. White, Science {\bf 332}, 1173 (2011).
\bibitem{depenbrock2012} S. Depenbrock, I.P. McCulloch, and U. Schollw\"ock,\prl {\bf  109}, 067201 (2012).
\bibitem{jiang2012} H.C. Jiang, Z.H. Wang, and L. Balents, Nat. Phys. {\bf 8}, 902 (2012).
\bibitem{clark2013} B. K. Clark, J. M. Kinder, E. Neuscamman, G. K.-L. Chan, and M. J. Lawler, \prl {\bf 111}, 187205 (2013).
\bibitem{liao2017} H.J. Liao, Z.Y. Xie, J. Chen, Z.Y. Liu, H.D. Xie, R.Z. Huang, B. Normand, and T. Xiang, \prl {\bf 118}, 137202 (2017).
\bibitem{hering2019} M. Hering, J. Sonnenschein, Y. Iqbal, and J. Reuther, \prb {\bf 99}, 100405(R) (2019).
\bibitem{becca2017} F. Becca and S. Sorella, {\it Quantum Monte Carlo Approaches for Correlated Systems} (Cambridge University Press, 2017).
\bibitem{affleck1988c} I. Affleck and J.B. Marston, \prb {\bf 37}, 3774(R) (1988).
\bibitem{ran2007} Y. Ran, M. Hermele, P.A. Lee, and X.-G. Wen, \prl {\bf 98}, 117205 (2007).
\bibitem{li2010} T. Li and F. Yang, \prb {\bf 81}, 214509 (2010).
\bibitem{ferrari2018} F. Ferrari, A. Parola, S. Sorella, and F. Becca, \prb {\bf 97}, 235103 (2018).
\bibitem{lieb1994} E.H. Lieb, \prl {\bf 73}, 2158 (1994).
\bibitem{ferrari2018b} F. Ferrari and F. Becca, \prb {\bf 98}, 100405 (2018).
\bibitem{dallapiazza2015} B. Dalla Piazza, M. Mourigal, N.B. Christensen, G.J. Nilsen, P. Tregenna-Piggott, T.G. Perring, M. Enderle, 
   D.F. McMorrow, D.A. Ivanov, and H.M. R\o{}nnow, Nat. Phys. {\bf 11}, 62 (2015).
\bibitem{zhang2020} C. Zhang and T. Li, \prb {\bf 102}, 195106 (2020).

\end{thebibliography}
\end{document}